\documentclass[a4paper,fleqn,usenatbib,useAMS]{mnras}

\usepackage{graphicx}  
\usepackage{amsmath}  
\usepackage{amssymb}  
\usepackage{multicol}        
\usepackage{bm}    
\usepackage{pdflscape}  
\usepackage{multirow}
\usepackage{xcolor}
\usepackage{hyperref}
\usepackage{booktabs}
\usepackage{subfigure}
\usepackage{xspace}
\usepackage{tikz}
\usepackage[para,online,flushleft]{threeparttable}
\usepackage[colorinlistoftodos,prependcaption,textsize=tiny]{todonotes}
\usepackage{xspace}
\usepackage{orcidlink}

\newcommand{\dd}{\mathrm d} 
\newcommand{\msun}{\mathrm M_\odot} 
\newcommand{\orcid}[1]{\href{https://orcid.org/#1}{\textcolor[HTML]{A6CE39}{\aiOrcid}}}

\def\specialname[#1]{\textbf{\textsc{#1}}}

\definecolor{lime}{HTML}{A6CE39}

\usepackage[T1]{fontenc}
\usepackage{ae,aecompl}
\usepackage{newtxtext,newtxmath}

\title[Merger-driven galaxy size growth and structural transformation]{
  Ruffled Feathers: Merger-driven galaxy size growth and structural transformation in EAGLE
}
\author[Wang et al.]{Kai Wang,$^{1,2}$\thanks{wkcosmology@gmail.com}\orcidlink{0000-0002-3775-0484}
  Carlton Baugh,$^{1,3}$\orcidlink{0000-0002-9935-9755}
  Sownak Bose,$^{1}$\orcidlink{0000-0002-0974-5266}
  Shaun Cole,$^{1}$\orcidlink{0000-0002-5954-7903}
  Carlos S. Frenk,$^{1}$\orcidlink{0000-0002-2338-716X}
  Cedric Lacey,$^{1}$\orcidlink{0000-0001-9016-5332}
  \newauthor
  Shengdong Lu,$^{1}$\orcidlink{0000-0002-6726-9499}
  Peder Norberg,$^{1,2}$\orcidlink{0000-0002-5875-0440}
  Isabel Santos-Santos,$^{4}$
  Francesco Shankar,$^{5}$\orcidlink{0000-0001-8973-5051}
  Tom Theuns$^{1}$\orcidlink{0000-0002-3790-9520}
  \newauthor
  and Tao Wang$^{6, 7}$\orcidlink{0000-0002-2504-2421}
  \\
  $^{1}$Institute for Computational Cosmology, Department of Physics, Durham University, South Road, Durham, DH1 3LE, UK\\
  $^{2}$Centre for Extragalactic Astronomy, Department of Physics, Durham University, South Road, Durham DH1 3LE, UK\\
  $^{3}$Institute for Data Science, Durham University, South Road, Durham DH1 3LE, UK\\
  $^{4}$Leibniz-Institut für Astrophysik Potsdam (AIP), An der Sternwarte 16, 14482 Potsdam, Germany\\
  $^{5}$School of Physics and Astronomy, University of Southampton, Highfield, Southampton SO17 1BJ, UK\\
  $^{6}$School of Astronomy and Space Science, Nanjing University, Nanjing, Jiangsu 210093, China\\
  $^{7}$Key Laboratory of Modern Astronomy and Astrophysics, Nanjing University, Ministry of Education, Nanjing 210093, China\\
}

\date{Last updated 2025 May 22; in original form 2025 May 5}

\pubyear{2025}

\begin{document}
\raggedbottom
\label{firstpage}
\pagerange{\pageref{firstpage}--\pageref{lastpage}}
\maketitle


\begin{abstract}
  Galaxy mergers drive both the size growth and the transformation from discs to spheroids, yet the prescriptions used to model these processes in semi-analytic frameworks have not been tested against the realistic merger population in cosmological hydrodynamical simulations.
  Using $\approx 4{,}500$ mergers identified in the EAGLE simulation, we test an energy-conservation estimator for post-merger galaxy sizes and quantify merger-driven morphological transformation.
  The predicted remnant half-stellar-mass radius matches the simulated descendant size with a scatter of $\approx 0.12$--$0.15$~dex and no significant systematic dependence on progenitor properties, while a commonly used dissipation correction applied to gas-rich mergers under-predicts the post-merger size by up to $\approx 0.4$~dex in a cosmological context and increases the overall scatter.
  The per-merger size growth increases monotonically with the stellar mass ratio of the merging pair, from $\lesssim 0.03$~dex for minor mergers to $\approx 0.10$~dex for equal-mass mergers.
  From the energy-conservation estimator, we analytically derive the size growth efficiency per unit accreted stellar mass, $\eta \equiv \mathrm{d}\log_{10} r_{\star}/\mathrm{d}\log_{10} M_{\star}$, and show that $\eta$ reaches $\approx 2$ only in the idealised limit of collisionless minor mergers with zero orbital energy; as $\eta$ is highly sensitive to the orbital energy at the time of merging, the minor merger channel cannot be established as the driver of the rapid size growth of massive galaxies without better constraints on this quantity.
  Beyond the size growth, mergers systematically reduce rotational support and increase triaxiality in proportion to mass ratio, but even the most nearly equal-mass mergers do not always fully destroy the disc, in tension with the complete disc destruction assumed in several semi-analytic models.
\end{abstract}


\section{Introduction}%
\label{sec:introduction}

The morphology and kinematics of galaxies are encoded in their distribution function in six-dimensional phase space, however, galaxies occupy a restricted set of structural and dynamical states.
The Hubble sequence captures this regularity, spanning rotation-supported discs and dispersion-supported spheroids.
Explaining how galaxies evolve between these states is central to understanding galaxy evolution.

In the $\Lambda$CDM cosmology, galaxies form in dark matter haloes that acquire angular momentum through tidal torques \citep{peeblesOriginAngularMomentum1969, catelanEvolutionAngularMomentum1996}.
Cooling gas can retain much of this angular momentum and settle into a rotationally supported disc \citep{peeblesOriginAngularMomentum1969, fallFormationRotationDisc1980, efstathiouStabilityMassesDisc1982, moFormationGalacticDiscs1998}.
This picture reproduces the observed mass--size relation and its scatter in the local Universe \citep[e.g.][]{shenSizeDistributionGalaxies2003, kravtsovSizeVirialRadiusRelation2013, wangDearthDifferencesCentral2020} and provides a viable framework at higher redshift \citep[e.g.][]{somervilleExplanationObservedWeak2008, somervilleRelationshipGalaxyDark2018}.
This framework predicts that galaxy size should correlate with halo spin, but hydrodynamical simulations have not reached a consensus on the strength of this correlation \citep[e.g.][]{rodriguez-gomezRoleMergersHalo2017, jiangDarkmatterHaloSpin2019, yangGalaxySizeHalo2023, liangConnectionGalaxyMorphology2025, sunControlledExperimentsDarkMatter2026}.

Several mechanisms can redistribute angular momentum and heat stellar orbits, including internal dynamical instabilities \citep{efstathiouStabilityMassesDisc1982}, environmental processes \citep{dresslerStellarDynamicsNuclei1988, mooreGalaxyHarassmentEvolution1996}, and galaxy mergers \citep{toomreGalacticBridgesTails1972, toomreMergersConsequences1977, hernquistStructureMergerRemnants1992, wangforgedquenching}.
Among these, mergers are particularly important: \citet{toomreMergersConsequences1977} first proposed that major mergers between spiral galaxies could produce elliptical galaxies, a hypothesis subsequently confirmed by numerical simulations \citep[e.g.][]{barnesEncountersDiskHalo1988, barnesDynamicsInteractingGalaxies1992, hernquistStructureMergerRemnants1992, barnesTransformationsGalaxiesII1996}.

Because merger-driven processes are highly nonlinear, numerical simulations are required to predict their outcomes, but simple physical arguments have also yielded widely used prescriptions.
A key example is the merger remnant size estimator introduced by \citet{coleHierarchicalGalaxyFormation2000}, which approximates the remnant size by enforcing energy conservation between the progenitors' binding energies and that of the descendant galaxy.
Idealised merger simulations showed that this estimator works well, while taking into account the dissipative radiative loss can further improve the accuracy of the predictor \citep{covingtonPredictingPropertiesRemnants2008}.
This prescription has been invoked to interpret the size growth of massive quiescent galaxies and the build-up of giant ellipticals \citep[e.g.][]{naabMinorMergersSize2009, bezansonRelationCompactQuiescent2009, hopkinsDiscriminatingPhysicalProcesses2010}, and is widely used to predict post-merger sizes in semi-analytic models \citep[e.g.][]{guoDwarfSpheroidalsCD2011, bensonALACTICUSSemianalyticModel2012, shankarSizeEvolutionSpheroids2013, shankarEnvironmentalDependenceBulgedominated2014, shankarAVOIDINGPROGENITORBIAS2015, xieSizeEvolutionElliptical2015, lagosSharkIntroducingOpen2018}.
However, the validity of the size estimator has not been tested against the realistic merger population of a cosmological hydrodynamical simulation, where progenitor structure and orbital parameters are drawn from a realistic distribution set by hierarchical growth rather than being prescribed by hand.

Here, we use the EAGLE simulation \citep{schayeEAGLEProjectSimulating2015, crainEAGLESimulationsGalaxy2015, mcalpineEagleSimulationsGalaxy2016}, which reproduces the observed galaxy mass--size relation at $z=0$ by calibration \citep{crainEAGLESimulationsGalaxy2015, furlongSizeEvolutionNormal2017} and, as a prediction, the observed morphological mix \citep{correaRelationGalaxyMorphology2017, thobRelationshipMorphologyKinematics2019}, to test these merger-driven predictions from semi-analytic prescriptions.
We identify merger events by following the merger tree that describes the hierarchical growth of a galaxy and evaluate the performance of the energy-conservation post-merger galaxy size estimator proposed by \citet{coleHierarchicalGalaxyFormation2000}.
We then measure how mergers change the morphological and kinematic properties of galaxies using a set of diagnostics from \citet{thobRelationshipMorphologyKinematics2019}.

This paper is organised as follows. Section~\ref{sec:simulation_data} describes the simulation, morphological and kinematic measurements, and the identification of galaxy mergers.
Section~\ref{sec:predicting_the_size_of_merger_remnant} tests the post-merger size estimator and quantifies the role of gas dissipation.
Section~\ref{sec:merger_driven_galaxy_size_growth} derives the size growth efficiency and its observational implications.
Section~\ref{sec:morphology_transformation_after_merger} quantifies merger-driven morphological transformation.
Section~\ref{sec:summary} summarises our conclusions.

\section{Simulation data}
\label{sec:simulation_data}

\begin{figure}
  \begin{center}
    \includegraphics[width=0.95\linewidth]{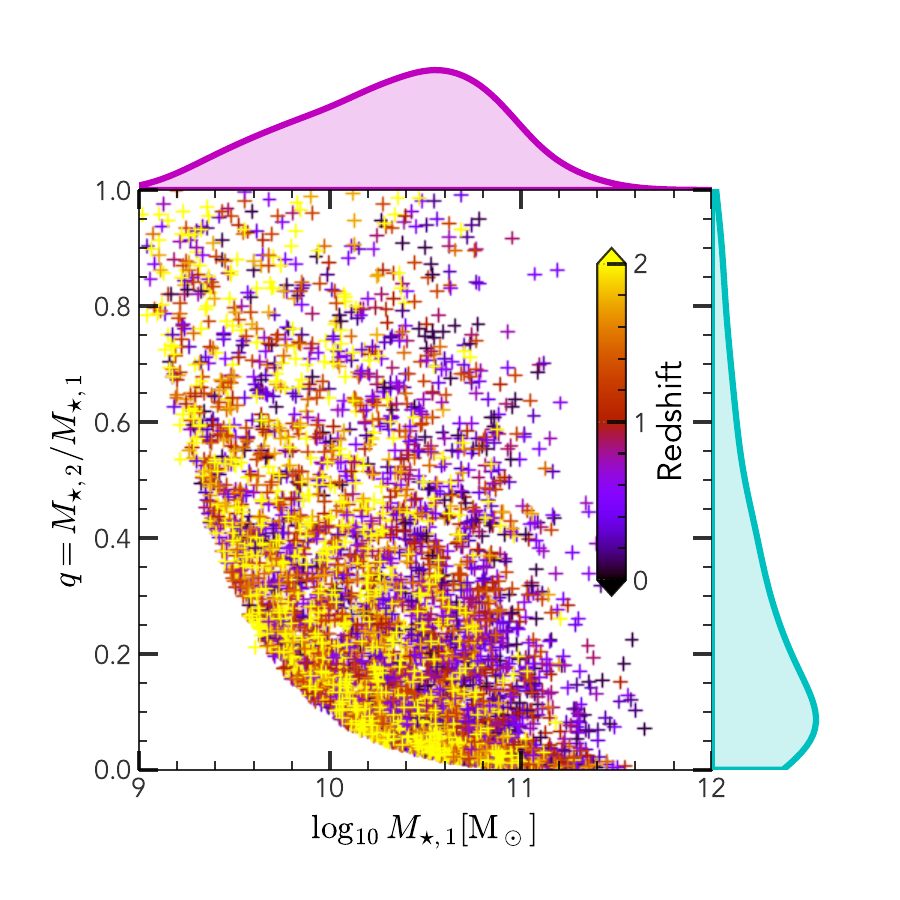}
  \end{center}
  \caption{
    The joint distribution of stellar mass ratio of merging progenitors, $q=M_{\star, 2}/M_{\star, 1}$, and the main progenitor stellar mass, $\log_{10} M_{\star, 1}$.
    Colours indicate the redshift of the merger.
    The shaded histograms show the marginal distributions of $q$ and $\log_{10} M_{\star, 1}$.
  }
  \label{fig:merger_distribution}
\end{figure}

We use the EAGLE cosmological hydrodynamical simulations \citep{schayeEAGLEProjectSimulating2015, crainEAGLESimulationsGalaxy2015}, focusing on the standard-resolution run \texttt{RefL100N1504}.
This simulation follows a periodic volume of comoving side length $L=100$ cMpc with $2\times 1504^3$ particles.
The initial baryonic and dark matter particle masses are $m_{\rm g}=1.8\times 10^6\,\msun$ and $m_{\rm dm}=9.70\times 10^6\,\msun$, respectively.
The Plummer-equivalent gravitational softening length is fixed in comoving coordinates to 2.66 ckpc for $z>2.8$ and to proper length of 0.7 pkpc thereafter.

Haloes are identified using a friends-of-friends algorithm and gravitationally bound substructures are found with \textsc{SUBFIND} \citep{springelPopulatingClusterGalaxies2001}.
We define the galaxy centre as the position of the particle of either type with the minimum gravitational potential of the host subhalo.
Unless stated otherwise, galaxy properties are computed from all stellar particles within a three-dimensional spherical aperture of radius 30 pkpc centred on the potential minimum.
Velocities are measured relative to the mean velocity of stellar particles within this aperture.
We denote the stellar half-mass radius measured within this aperture by $r_{\star}$.

We take the morphological and kinematical properties measured in \citet{thobRelationshipMorphologyKinematics2019}.
Stellar kinematics are measured within a 30 pkpc spherical aperture.
We define the rotation axis as the direction of the stellar angular momentum within this aperture, which defines the component of each particle's angular momentum along the rotation axis, which we take to be the $z$-axis, $L_{z, i}$.
The disc-to-total stellar mass ratio, $\rm D/T$, is obtained using the standard estimator in which the bulge mass equals twice the mass in counter-rotating stellar particles, i.e. those particles with $L_{z, i} < 0$,
\begin{equation}
  {\rm D/T} = 1 - \frac{2}{\sum_{i}m_i}\sum_{i, L_{z, i} < 0}m_i ,
\end{equation}
where $m_i$ is the mass for the $i$-th stellar particle.

We also use the fraction of kinetic energy invested in ordered co-rotation, $\kappa_{\rm co}$, \citep{thobRelationshipMorphologyKinematics2019}
\begin{equation}
  \kappa_{\rm co}
  = \frac{K_{\rm co}^{\rm rot}}{K}
  = \frac{1}{K}\sum_{i,L_{z,i}>0}\frac{1}{2}m_i
  \left(\frac{L_{z,i}}{m_i r_i}\right)^2,
\end{equation}
where $r_i$ is the cylindrical radius in the plane perpendicular to the rotation axis, $K=\sum_i\tfrac{1}{2}m_iv_i^2$ is the total kinetic energy in the centre-of-mass frame, and $v_i$ is the magnitude of the particle velocity relative to the galaxy centre-of-mass velocity.

Finally, galaxy morphology is quantified by fitting an ellipsoid to the stellar mass distribution using an iterative reduced inertia tensor\footnote{The reduced inertia tensor is defined as
  \begin{equation*}
    \mathcal{M}_{ij} = \frac{\sum_p m_p \, r_{p,i} \, r_{p,j} / r_p^2}
    {\sum_p m_p / r_p^2},
  \end{equation*}
  where the sum runs over all bound stellar particles within 30~pkpc, $m_p$ is the particle mass, $r_p$ is its distance from the galaxy centre, and $r_{p,i}$ with $i=1,2,3$ is the $i$-th component of the particle position vector.
  The $r_p^{-2}$ weighting suppresses the contribution of particles in the galaxy outskirts, making the shape measurement more sensitive to the inner structure.}.
The iteration begins from the set of stellar particles within a spherical aperture of 30~pkpc and converges when both $b/a$ and $c/a$ change by less than 1~per~cent between successive iterations, with the ellipsoid volume held fixed throughout \citep[see][for details]{thobRelationshipMorphologyKinematics2019}.
The axis lengths $a\ge b\ge c$ are obtained from the eigenvalues of the converged tensor and define the flattening and triaxiality,
\begin{equation}
  \epsilon \equiv 1 - \frac{c}{a}, \qquad T\equiv \frac{a^2 - b^2}{a^2 - c^2} .
\end{equation}
Here $\epsilon=0$ corresponds to a sphere, and low (high) values of $T$ correspond to oblate (prolate) ellipsoids.

We identify galaxy mergers from the merger trees by linking galaxies between consecutive snapshots via their unique descendant.
A merger occurs when two galaxies in snapshot $s$ share the same descendant in snapshot $s+1$.
If more than two progenitors share a descendant, we keep the two most massive progenitors and discard the rest.
We label the more massive progenitor as Progenitor-1 (the main progenitor) and the less massive one as Progenitor-2 (the secondary progenitor).
We append suffixes 1 and 2 to all progenitor properties.
All progenitor properties are measured at the last snapshot before the merger, and all descendant properties at the first snapshot after.
We restrict our analysis to galaxies with $M_{\star} > 10^9\,\rm M_\odot$, the mass range over which EAGLE reproduces the observed galaxy mass--size relation \citep{crainEAGLESimulationsGalaxy2015, furlongSizeEvolutionNormal2017}.

Fig.\,\ref{fig:merger_distribution} shows the joint distribution of the main progenitor stellar mass, $M_{\star, 1}$, and the progenitor stellar mass ratio
\begin{equation}
  q\equiv M_{\star, 2}/M_{\star, 1},
\end{equation}
together with the marginal distributions.
We identify 4,547 merger events.
Of these, about 31 per cent have $q < 0.1$, while 33, 25, and 11 per cent lie in $0.1 \leq q < 0.3$, $0.3 \leq q < 0.6$, and $0.6 \leq q < 1$, respectively.
The lack of mergers at low $M_{\star, 1}$ and low $q$ partly reflects our requirement that both progenitor galaxies satisfy $M_{\star} > 10^9\,\msun$.

The colour-coding of merger redshift indicates that the merger rate in our sample is dominated by low redshift events.
Approximately half of the mergers occur at $z < 1$, about 80 per cent at $z<2$, and about 95 per cent at $z<3$.
In addition, about 80 per cent of the mergers lead to a central descendant galaxy.
Satellite--satellite mergers therefore contribute only a minority of the merger population in our selection.

\section{Predicting the size of merger remnant}
\label{sec:predicting_the_size_of_merger_remnant} 

\subsection{Post-merger galaxy size estimate based on energy conservation}
\label{sub:post_merger_galaxy_size_estimate_based_on_energy_conservation} 

\begin{figure*}
  \begin{center}
    \includegraphics[width=0.95\linewidth]{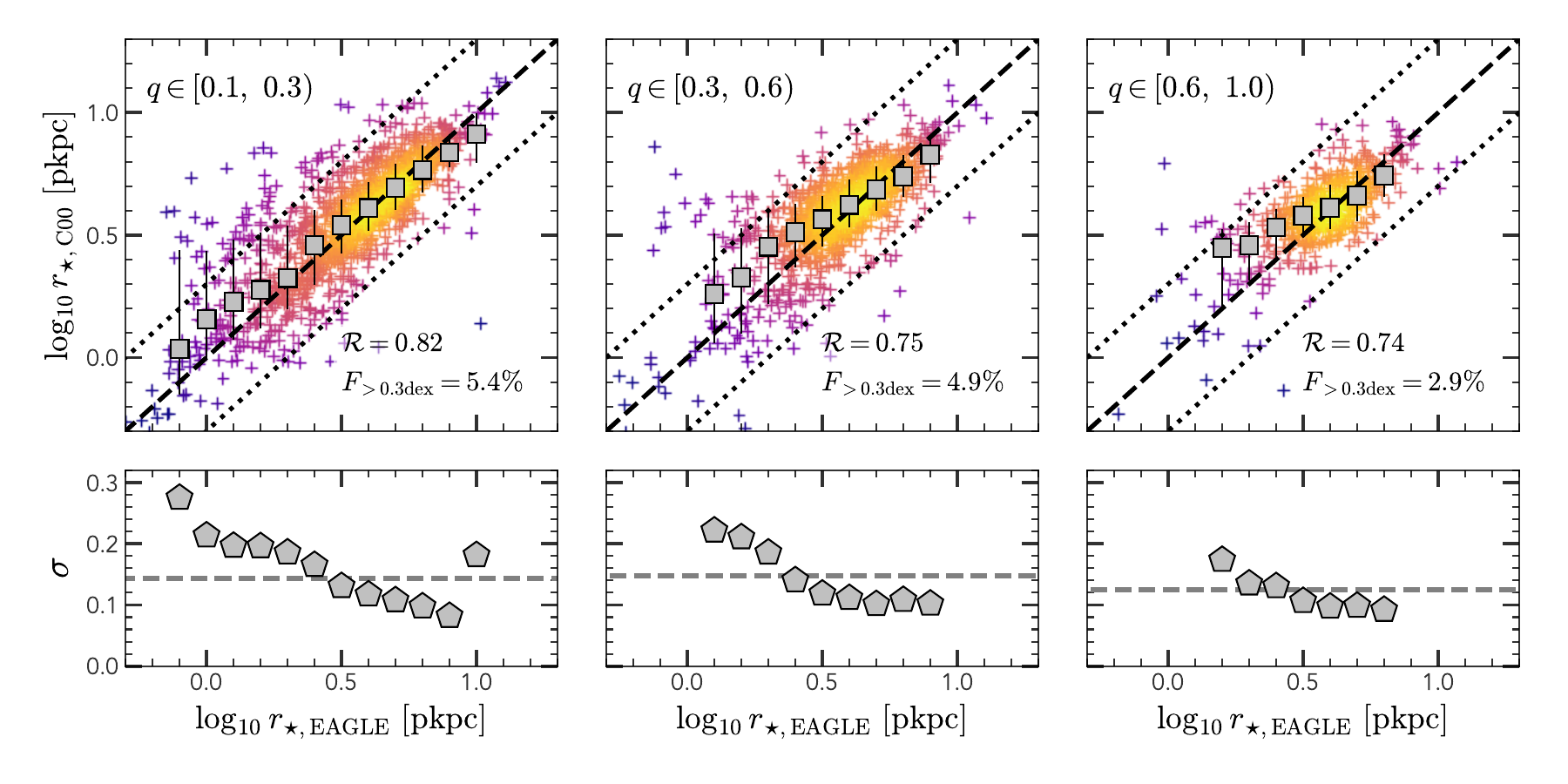}
  \end{center}
  \caption{
    Galaxy size of merger remnants predicted by the energy-conservation estimator in \eqref{eq:merger_size} with $f_{\rm orbit}/c=2$ compared to the descendant galaxy size in EAGLE.
    Each panel shows a different merger mass-ratio bin, $q\equiv M_{\star, 2}/M_{\star, 1}$: $0.1\leq q < 0.3$ (left), $0.3\leq q < 0.6$ (middle), $0.6 \leq q < 1$ (right).
    Symbols show individual mergers, coloured by the local density of symbols in 0.1-dex bins.
    Grey squares give the median $\log_{10}(r_{\star, \rm C00}/{\rm pkpc})$ in bins of $\log_{10}(r_{\star, \rm EAGLE}/{\rm pkpc})$ with a bin width of 0.2 dex.
    Error bars indicate the 16th-84th percentiles.
    The dashed line marks equality, and the dotted lines show $\pm 0.3$ dex.
    Each panel lists Spearman's rank correlation coefficient and the fraction of outliers with $|\log_{10}(r_{\star, \rm C00}/r_{\star, \rm EAGLE})| > 0.3$ dex.
    The lower panels show the standard deviation, $\sigma$, of $\log_{10}(r_{\star, \rm C00}/r_{\star, \rm EAGLE})$ in bins of $\log_{10}(r_{\star, \rm EAGLE}/{\rm pkpc})$.
    The horizontal dashed line marks the standard deviation of $\log_{10}(r_{\star,\rm C00}/r_{\star,\rm EAGLE})$ for all merger events in each mass-ratio bin, providing a reference for the overall scatter independent of $r_{\star,\rm EAGLE}$.
    The estimator correlates well with the simulated remnant sizes in all merger-ratio bins ($\mathcal R \approx 0.8$) with a small outlier fraction ($\lesssim 5$ per cent).
    The overall residual scatter is $\sigma\approx 0.12$--$0.15$ dex.
  }
  \label{fig:merger_size}
\end{figure*}

\citet{coleHierarchicalGalaxyFormation2000} introduced a simple estimate for the size of the merger remnant that has since been adopted in many semi-analytic models \citep[e.g.][]{guoDwarfSpheroidalsCD2011, covingtonRoleDissipationScaling2011, porterUnderstandingStructuralScaling2014, xieSizeEvolutionElliptical2015, lagosSharkIntroducingOpen2018}.
The estimate assumes that the internal binding energies of the progenitors and their orbital energy combine and are conserved during the merger, with the remnant relaxing to virial equilibrium.
With the total energy\footnote{The original prescription in \citet{coleHierarchicalGalaxyFormation2000} includes all mass associated with the galaxy, including gas and dark matter. We retain only the stellar mass and discuss this choice later in this section.} of a virialised system as $E=-cGM_{\star}^2/r_{\star}$, where $r_{\star}$ is the characteristic size and $c$ encapsulates the density profile (typically $c\approx 0.5$), energy conservation gives
\begin{equation}
  \frac{(M_{\star, 1} + M_{\star, 2})^2}{r_{\star,\rm C00}} = \frac{M_{\star, 1}^2}{r_{\star, 1}} + \frac{M_{\star, 2}^2}{r_{\star, 2}} + \frac{f_{\rm orbit}}{c}\frac{M_{\star, 1}M_{\star, 2}}{r_{\star, 1} + r_{\star, 2}} ,
  \label{eq:merger_size}
\end{equation}
where $M_{\star, 1}$ and $M_{\star, 2}$ are the progenitor masses and $r_{\star, 1}$ and $r_{\star, 2}$ are their corresponding sizes.
On the right side of the equation, the first two terms represent the progenitors' self-binding energies, and the final term is approximately the orbital contribution at the time the galaxies merge, with $f_{\rm orbit}$ parameterising the orbital energy.
For $f_{\rm orbit}=1$, this term equals the energy of two point masses on a circular orbit with separation $r_{\star, 1} + r_{\star, 2}$ \citep[see also][]{coleHierarchicalGalaxyFormation2000}.
In equation~\eqref{eq:merger_size}, $r_{\star,\rm C00}$ is the predicted post-merger size, which depends on the choice of $f_{\rm orbit}/c$; we adopt $f_{\rm orbit}/c = 2$ following \citet{coleHierarchicalGalaxyFormation2000}, unless stated otherwise.
Appendix~\ref{sec:Optimising_forbit} examines the sensitivity of the estimator to this choice.

We estimate the remnant size, $r_{\star,\rm C00}$, from equation~\eqref{eq:merger_size} and compare it to the descendant galaxy size measured in the simulation, $r_{\star,\rm EAGLE}$.
Fig.~\ref{fig:merger_size} shows the results in three bins of the merger mass ratio.
Each point corresponds to a merger event; the colour scale encodes the local point density evaluated in 0.1-dex bins.
Grey squares show the median $\log_{10}(r_{\star,\rm C00}/{\rm pkpc})$ in bins of $\log_{10}(r_{\star,\rm EAGLE}/{\rm pkpc})$, with error bars indicating the 16th-84th percentiles.
Each panel shows Spearman's rank correlation coefficient and the fraction of outliers, defined as $|\log_{10} (r_{\star,\rm C00}/r_{\star,\rm EAGLE} ) | > 0.3\,\rm dex$.
The lower panels show the standard deviation of $\log_{10}(r_{\star,\rm C00}/r_{\star,\rm EAGLE})$ as a function of $r_{\star,\rm EAGLE}$ with the horizontal line indicating the residual standard deviation in each merger ratio bin.

As shown in Fig.~\ref{fig:merger_size}, the energy-conservation size estimator reproduces the sizes of merger remnants in EAGLE.
The predicted and simulated half-stellar-mass radii correlate strongly in all three merger-ratio bins, with Spearman's rank correlation coefficient $\mathcal R\approx 0.8$.
The binned medians lie close to the one-to-one relation over the full dynamical range, indicating no systematic bias in the mean relation.
The residual has a standard deviation of about $0.12$--$0.15$ dex.
Appendix~\ref{sec:size_residual_dependence} shows that the residual is insensitive to the main progenitor stellar mass, morphology, kinematics, and redshift, with the exception of an over-prediction of $\approx 0.1$ dex for gas-rich minor mergers.
Appendix~\ref{sec:info_gain} quantifies the improvement in predictive accuracy gained by including the secondary progenitor size relative to the null assumption that the descendant size equals that of the main progenitor.

The original prescription of \citet{coleHierarchicalGalaxyFormation2000} uses the total mass associated with each galaxy, including gas and dark matter, though several semi-analytic models evaluate it using stellar mass alone \citep[e.g.][]{naabMinorMergersSize2009, guoDwarfSpheroidalsCD2011, henriquesGalaxyFormationPlanck2015}.
We follow the latter approach and find that it already reproduces the EAGLE merger remnant sizes without systematic bias.
Since including the dark matter mass is known to increase the predicted post-merger size \citep{gonzalezTestingModelPredictions2009, shankarSizeEvolutionSpheroids2013}, doing so would likely over-predict the simulated descendant sizes given that the stellar-mass-only estimator is already unbiased.

\subsection{The role of gas dissipation in merger remnant sizes}
\label{sub:the_role_of_gas_in_galaxy_merger}

\begin{figure*}
  \begin{center}
    \includegraphics[width=0.8\linewidth]{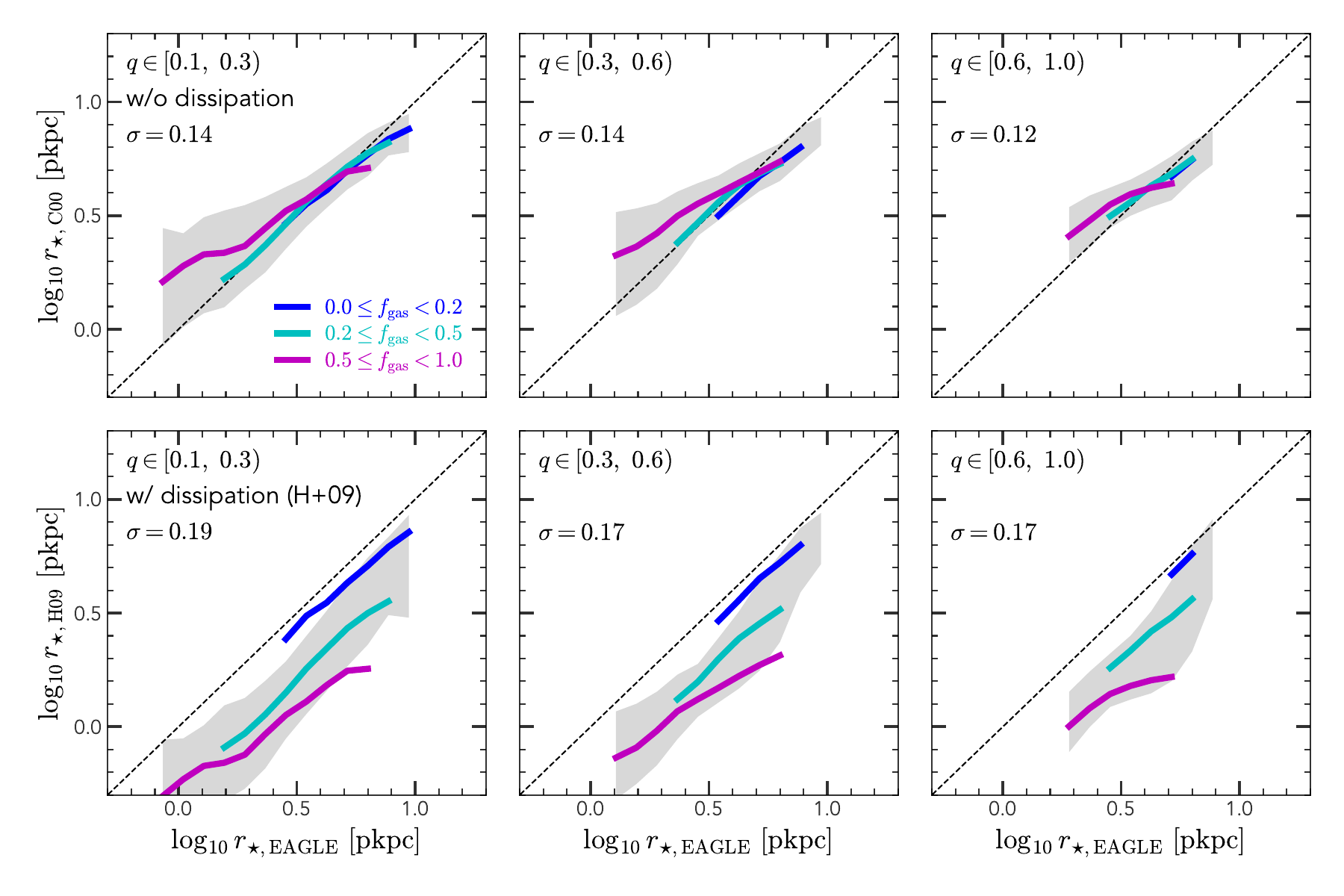}
  \end{center}
  \caption{
    Post-merger galaxy size estimates compared to the simulated descendant size in EAGLE, in bins of merger mass ratio (columns) and progenitor gas fraction $f_{\rm gas}$ (colours).
    The top and bottom rows show the energy-conservation estimator of \citet{coleHierarchicalGalaxyFormation2000} without dissipation correction, evaluated with $f_{\rm orbit}/c = 2$, and the dissipation-corrected estimator of \citet{hopkinsDissipationExtraLight2009}, evaluated with $f_{\rm orbit}/c = 0$ following their recommendation.
    In each panel, the grey shaded region shows the 16th--84th percentile range of the full population, and coloured solid lines show the median predicted size in bins of $r_{\star, \rm EAGLE}$ for three gas fraction bins: $0.0 \leq f_{\rm gas} < 0.2$, $0.2 \leq f_{\rm gas} < 0.5$, and $0.5 \leq f_{\rm gas} < 1.0$.
    The residual scatter $\sigma$ of the full population is indicated in each panel.
    The \citet{coleHierarchicalGalaxyFormation2000} estimator performs well across all mass-ratio and gas-fraction bins, with a modest over-prediction for compact remnants.
    The dissipation correction of \citet{hopkinsDissipationExtraLight2009} over-corrects for gas-rich galaxies, under-predicting their sizes by up to $\approx 0.4$~dex and increasing the overall scatter.
  }
  \label{fig:merger_size_fgasbin}
\end{figure*}

\citet{covingtonPredictingPropertiesRemnants2008} showed, using idealised merger simulations, that the energy-conservation estimator of \citet{coleHierarchicalGalaxyFormation2000} over-predicts post-merger galaxy sizes because it neglects dissipation during the merger.
They proposed a refined estimator to account for this effect.
Assuming parabolic orbits and progenitor galaxies with characteristic structural properties, \citet{hopkinsDissipationExtraLight2009} showed that the dissipation correction of \citet{covingtonPredictingPropertiesRemnants2008} reduces to a simple multiplicative factor,
\begin{equation}
  r_{\star, \rm H09} = \frac{r_{\star, \rm C00}(f_{\rm orbit}/c = 0)}{1 + f_{\rm gas}/f_0},
  \label{eq:merger_size_H09}
\end{equation}
where $f_{\rm gas}$ is the ratio of the total gas mass to the total gas plus stellar mass of the two progenitors,\footnote{The original prescription also includes stars formed during the merger-induced starburst in the denominator of $f_{\rm gas}$.
  We do not model this process explicitly; replacing it with the total stellar and gas mass of the descendant galaxy leaves the results essentially unchanged.} and $f_0 \approx 0.25$ \citep{hopkinsDissipationExtraLight2009, shankarSizeEvolutionSpheroids2013}.

Fig.~\ref{fig:merger_size_fgasbin} compares these two estimators in bins of gas fraction.
The top panels confirm that the original \citet{coleHierarchicalGalaxyFormation2000} estimator slightly over-predicts the size of compact remnants, though the overall scatter is $\sigma \approx 0.12$--$0.15$~dex.
The bottom panels show that the dissipation correction of \citet{hopkinsDissipationExtraLight2009} over-corrects for gas-rich mergers, under-predicting their sizes by up to $\approx 0.4$~dex and increasing the overall scatter to $\sigma \approx 0.17$--$0.19$~dex.
This failure in a cosmological context is unsurprising: the correction was calibrated on idealised simulations that lack a realistic prior on pre-merger configurations.
The difference before and after correction is consistent with \citet{shankarSizeEvolutionSpheroids2013}, who find that this correction reduces predicted sizes of early-type galaxies near $M_\star\sim 10^{10}~{\rm M}_\odot$ by $\approx 0.5$~dex.
We attempted a recalibration of the dissipation correction against the EAGLE residuals, but the improvement over the original estimator is marginal.
We therefore retain the \citet{coleHierarchicalGalaxyFormation2000} estimator throughout, treating the gas-fraction dependence of its residuals as a subdominant systematic.

\section{Merger-driven galaxy size growth}
\label{sec:merger_driven_galaxy_size_growth}

\begin{figure}
  \begin{center}
    \includegraphics[width=0.95\linewidth]{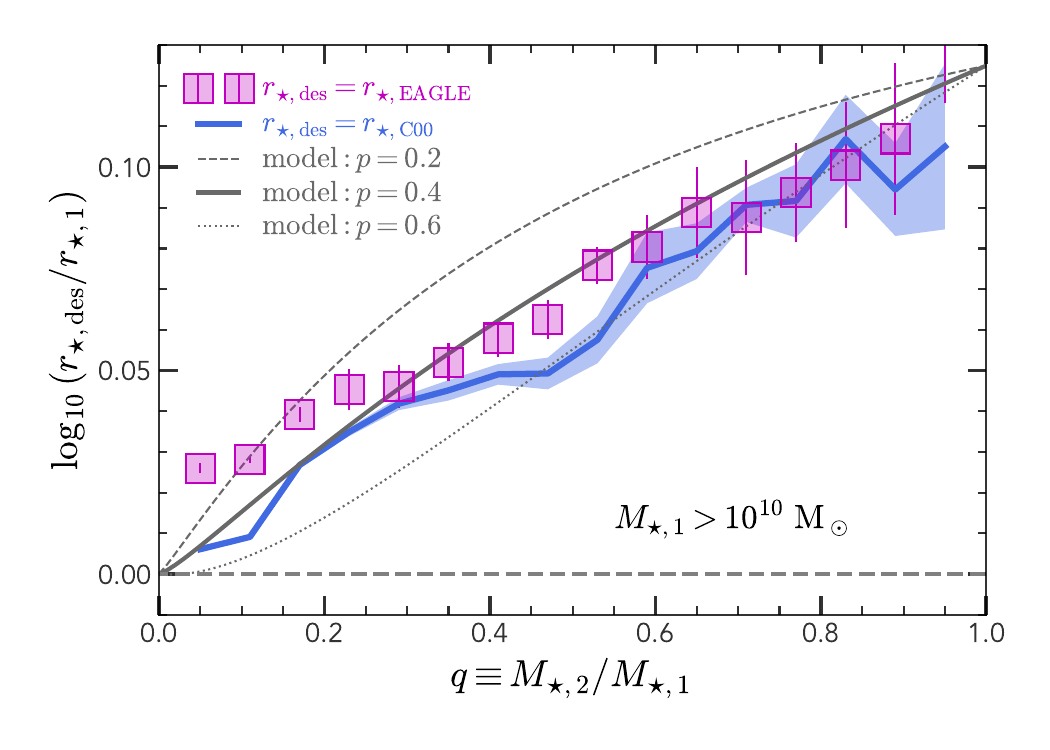}
  \end{center}
  \caption{
    Post-merger size growth as a function of merger stellar mass ratio $q \equiv M_{\star,2}/M_{\star,1}$, for mergers with the main progenitor stellar mass $M_{\star,1} > 10^{10}\,\rm M_{\odot}$, a threshold imposed to ensure a complete sample across the full range of mass ratios $q \geq 0.1$.
    The $y$-axis shows the logarithmic size growth of the descendant relative to the main progenitor, $\log_{10}(r_{\star,\rm des}/r_{\star, 1})$.
    Magenta squares show the median size growth measured directly in EAGLE, with error bars indicating the uncertainty on the median estimated from 100 bootstrap realisations.
    The blue solid curve shows the median size growth predicted by the energy-conservation estimator of equation~\eqref{eq:merger_size}, with the shaded region indicating the bootstrap uncertainty.
    The dashed, solid, and dotted curves show results for mass--size relation slopes of $p=0.2$, $0.4$ and $0.6$, respectively.
    Both the EAGLE measurements and the energy-conservation estimator show a monotonically increasing size growth with merger mass ratio, and the two are in good agreement across the full range of $q$.
  }
  \label{fig:postmerger_size_inc}
\end{figure}

\begin{figure}
  \begin{center}
    \includegraphics[width=0.95\linewidth]{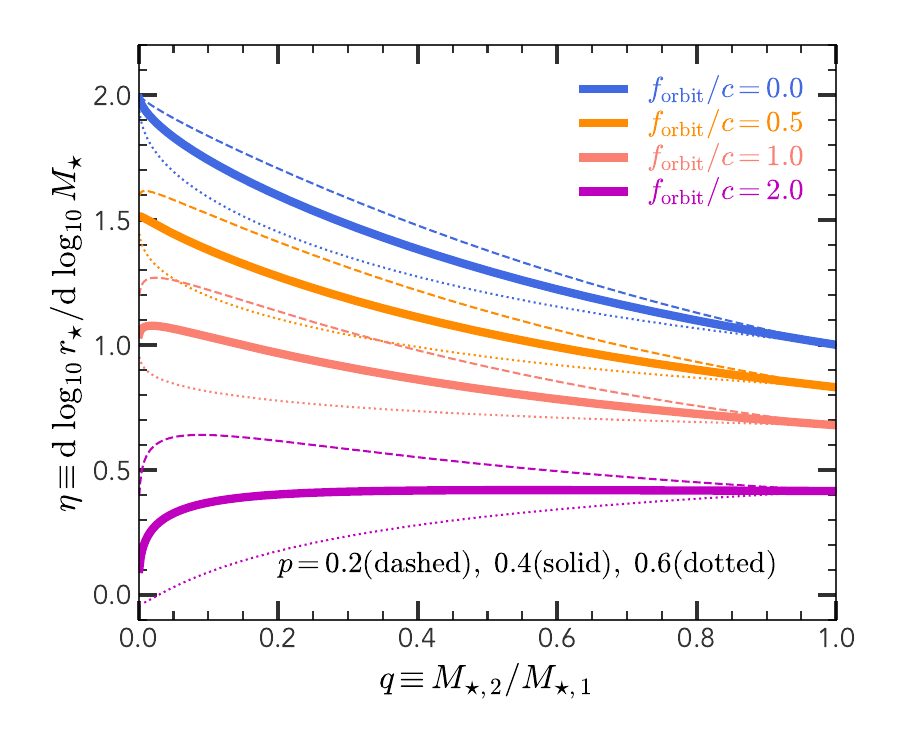}
  \end{center}
  \caption{
    The logarithmic derivative of galaxy size with respect to stellar mass, $\eta \equiv \mathrm{d}\log_{10} r_\star / \mathrm{d}\log_{10} M_\star$, as a function of merger stellar mass ratio, $q \equiv M_{\star,2}/M_{\star,1}$, predicted analytically from equation~\eqref{eq:eta} for four values of the orbital energy parameter $f_\mathrm{orbit}/c$: 0, 0.5, 1, and 2.
    The dashed, solid, and dotted curves show results for mass--size relation slopes of $p=0.2$, $0.4$ and $0.6$, respectively, illustrating the sensitivity of $\eta$ to the assumed slope of the progenitor mass--size relation.
    For equal-mass major mergers ($q\approx 1$), $\eta$ shows weak dependence on both $f_\mathrm{orbit}/c$ and $p$, with values ranging from approximately 0.4 to 1.
    In contrast, minor mergers ($q\rightarrow 0$) show a much stronger dependence on $f_\mathrm{orbit}/c$, with $\eta$ spanning the full range from 0 to 2.
    In general, larger values of $f_\mathrm{orbit}/c$ and steeper mass--size relation slopes $p$ both lead to weaker size growth per unit accreted mass across all merger mass ratios.
  }
  \label{fig:merger_driven_size}
\end{figure}

One of the most striking observational results in galaxy evolution is the dramatic size growth of massive galaxies since $z \sim 2$: at fixed stellar mass, galaxies at high redshift are a factor of several smaller than their local counterparts \citep[e.g.][]{daddiPassivelyEvolvingEarlyType2005, vandokkumConfirmationRemarkableCompactness2008, hillMassColorStructural2017, jiaSizeGrowthShort2024, songTransitionOutsideinInsideOut2026}.
Galaxy mergers are a leading candidate for driving this growth, as they add both stellar mass and kinetic energy to the remnant, expanding its effective radius \citep[e.g.][]{coleHierarchicalGalaxyFormation2000, naabMinorMergersSize2009}.
A single major merger can increase galaxy size by up to a factor of $\sim 2$, while a single minor merger has a negligible effect \citep[e.g.][]{naabMinorMergersSize2009, oserCosmologicalSizeVelocity2012, furlongSizeEvolutionNormal2017}.
However, in a cosmological context, galaxies undergo far more minor mergers than major ones over their lifetimes, and their cumulative contribution to size growth can be substantial, though major mergers still dominate the \textit{ex-situ} mass assembly \citep{rodriguez-gomezRoleMergersHalo2017}.
\citet{naabMinorMergersSize2009} proposed that this cumulative minor merger channel is a primary driver of the size growth of massive spheroidal galaxies \citep[see also][]{bezansonRelationCompactQuiescent2009, hopkinsDiscriminatingPhysicalProcesses2010}, but their treatment neglected the orbital energy contribution and assumed identical progenitor sizes, leaving the predicted growth rates uncertain at the quantitative level that is needed to confront observations.

Here, leveraging the cosmological prior on the merger population provided by EAGLE, we quantify the size growth \textit{per merger event} and its dependence on mass ratio.
We restrict this analysis to mergers with main progenitor stellar mass $M_{\star,1} > 10^{10}\,\rm M_{\odot}$ to ensure a complete sample across the full range of mass ratios $q \gtrsim 0.1$.
The magenta squares in Fig.~\ref{fig:postmerger_size_inc} show the median logarithmic size growth of the descendant relative to the main progenitor, $\log_{10}(r_{\star,\rm EAGLE}/r_{\star,1})$, as a function of merger mass ratio $q$.
The size growth increases monotonically with $q$, ranging from $\lesssim 0.03$~dex for minor mergers ($q \approx 0.1$) to $\sim 0.10$~dex for equal-mass mergers ($q \approx 1$).
The energy-conservation estimator ($r_{\star, \rm C00} / r_{\star, 1}$), shown in the blue solid curve and the shaded region, is in excellent agreement with the EAGLE measurements across the full range of $q$.

Rearranging equation~\eqref{eq:merger_size} and assuming that the progenitor galaxies follow a mass--size relation of the form $r_{\star} \propto M_{\star}^p$, the predicted size growth \textit{per merger event} can be written as (see Appendix~\ref{sec:derivation_size_growth} for a detailed derivation)
\begin{equation}
  \frac{r_{\star,\rm C00}}{r_{\star, 1}} = {(1 + q)^2}\left({1 + q^{2 - p} + \frac{f_{\rm orbit}}{c}\frac{q}{1 + q^p}}\right)^{-1}.
  \label{eq:size_growth}
\end{equation}
Adopting $p=0.4$ and $f_{\rm orbit}/c=2$, this expression reproduces the EAGLE result well, as shown by the grey solid curve in Fig.~\ref{fig:postmerger_size_inc}.
Appendix~\ref{sec:mass_size_relation_for_merging_progenitor_galaxies} shows that $p=0.4$ is a good description of the mass--size relation of the merger progenitor population in EAGLE.

To characterise the cumulative effect of mergers on galaxy size growth, we compute the logarithmic derivative of size with respect to accreted stellar mass (see Appendix~\ref{sec:derivation_size_growth} for a detailed derivation),
\begin{align}
  \eta(p, q, f_{\rm orbit}/c) & \equiv \frac{\dd \log_{10} r_{\star}}{\dd \log_{10}M_{\star}} = \frac{\log_{10}\left(r_{\star,\rm C00}/r_{\star, 1}\right)}{\log_{10}(1 + q)}\nonumber \\
                              & = 2 - \frac{\log_{10}\left(1 + q^{2 - p} + \frac{f_{\rm orbit}}{c}\frac{q}{1 + q^p}\right)}{\log_{10}(1 + q)}
  \label{eq:eta}
\end{align}
which measures the efficiency of size growth \textit{per unit logarithmic mass} accreted through mergers of a given mass ratio, and therefore governs how rapidly galaxies grow in size relative to their mass growth through mergers.
In the limits of vanishing and equal-mass mergers, this reduces, respectively, to\footnote{The minor merger limit assumes $p < 1$. For $p = 1$ the residual term $q^{1-p}$ in the expansion of equation~\eqref{eq:eta} is independent of $q$ and $\eta \rightarrow 1 - f_{\rm orbit}/c$ instead.}
\begin{align}
  \eta(p, q\rightarrow 0, f_{\rm orbit}/c) & \rightarrow 2 - \frac{f_{\rm orbit}}{c},                                               \\
  \eta(p, q\rightarrow 1, f_{\rm orbit}/c) & \rightarrow 2 - \frac{\log_{10}\left(2 + \frac{f_{\rm orbit}}{2c}\right)}{\log_{10}2}.
\end{align}
Fig.~\ref{fig:merger_driven_size} shows $\eta$ as a function of $q$ predicted by this analytical relation for four values of $f_{\rm orbit}/c=0, 0.5, 1, 2$ and three slopes $p =0.2, 0.4, 0.6$.
For equal-mass major mergers, $\eta$ lies in the range $0.4$--$1$ and depends weakly on both $f_{\rm orbit}/c$ and $p$.
For minor mergers, the dependence on $f_{\rm orbit}/c$ is much stronger, with $\eta$ spanning the full range from 0 to 2 as $f_{\rm orbit}/c$ decreases from 2 to 0, reflecting the increasing relative importance of the orbital energy term in the limit of very unequal mass ratios.
In general, larger $f_{\rm orbit}/c$ and steeper $p$ both reduce the efficiency of merger-driven size growth across all mass ratios.

The result that minor mergers can drive more efficient size growth per unit accreted mass than major mergers may appear to contradict \citet{nipotiGalaxyMergingFundamental2003}, who found that growing a galaxy through multiple equal-mass major mergers and through sequential accretion of small satellites, both on parabolic orbits ($f_\mathrm{orbit}/c = 0$), yields the same linear size growth rate per unit mass, i.e. $\eta =1$.
However, we note that their accretion hierarchy assigns satellite sizes consistent with $p = 1$ in our analytic framework.
As we show analytically here, when $p = 1$ and $f_\mathrm{orbit}/c = 0$, we have $\eta = 1$ exactly from equation~\eqref{eq:eta}, independently of the merger mass ratio $q$.
The equal size growth efficiency across all mass ratios in \citet{nipotiGalaxyMergingFundamental2003} is therefore a direct consequence of their progenitor size configuration rather than a general property of dissipationless merging.

The upper limit of $\eta \approx 2$ is only reached in the idealised case of purely collisionless minor mergers with no orbital energy contribution ($f_{\rm orbit}/c = 0$); any non-zero orbital energy or dissipative gas component will reduce $\eta$ below this value, and $\eta$ is highly sensitive to the orbital energy at the time of merging.
This has direct observational implications.
\citet{vandokkumGROWTHMASSIVEGALAXIES2010} used the \textit{constant} cumulative number density method to trace the mass and size growth of massive elliptical galaxies ($M_{\star} > 10^{11}~{\rm M}_\odot$) since $z \sim 2$ and inferred $\eta \approx 2$, implying very rapid size growth relative to mass growth.
Our results demonstrate that this value lies at the upper limit of what the energy-conservation argument predicts for merger-driven growth (equation~\ref{eq:eta}), attainable only under the most favourable conditions of purely minor, collisionless mergers with $f_{\rm orbit}/c = 0$ (i.e.\ parabolic orbits).
The refined \textit{evolving} cumulative number density method \citep[see also][]{behrooziUSINGCUMULATIVENUMBER2013, clauwensLargeDifferenceProgenitor2016, clauwensAverageStructuralEvolution2017a, wangRelatingGalaxiesDifferent2023} yields a more moderate estimate: \citet{hillMassColorStructural2017} inferred $\eta \approx 1.4$ from $z \sim 2$ to $0$ over the epoch in which mergers dominate mass growth.
Even this lower value requires a small orbital energy contribution ($f_{\rm orbit}/c \lesssim 0.5$) and assembly predominantly through minor mergers ($q \approx 0.1$), demonstrating that the minor merger channel cannot be established as the driver of the rapid size growth of massive galaxies without better constraints on the orbital energy at the time of merging.

\section{Merger-driven Morphology transformation}
\label{sec:morphology_transformation_after_merger} 

\begin{figure*}
  \begin{center}
    \includegraphics[width=0.95\linewidth]{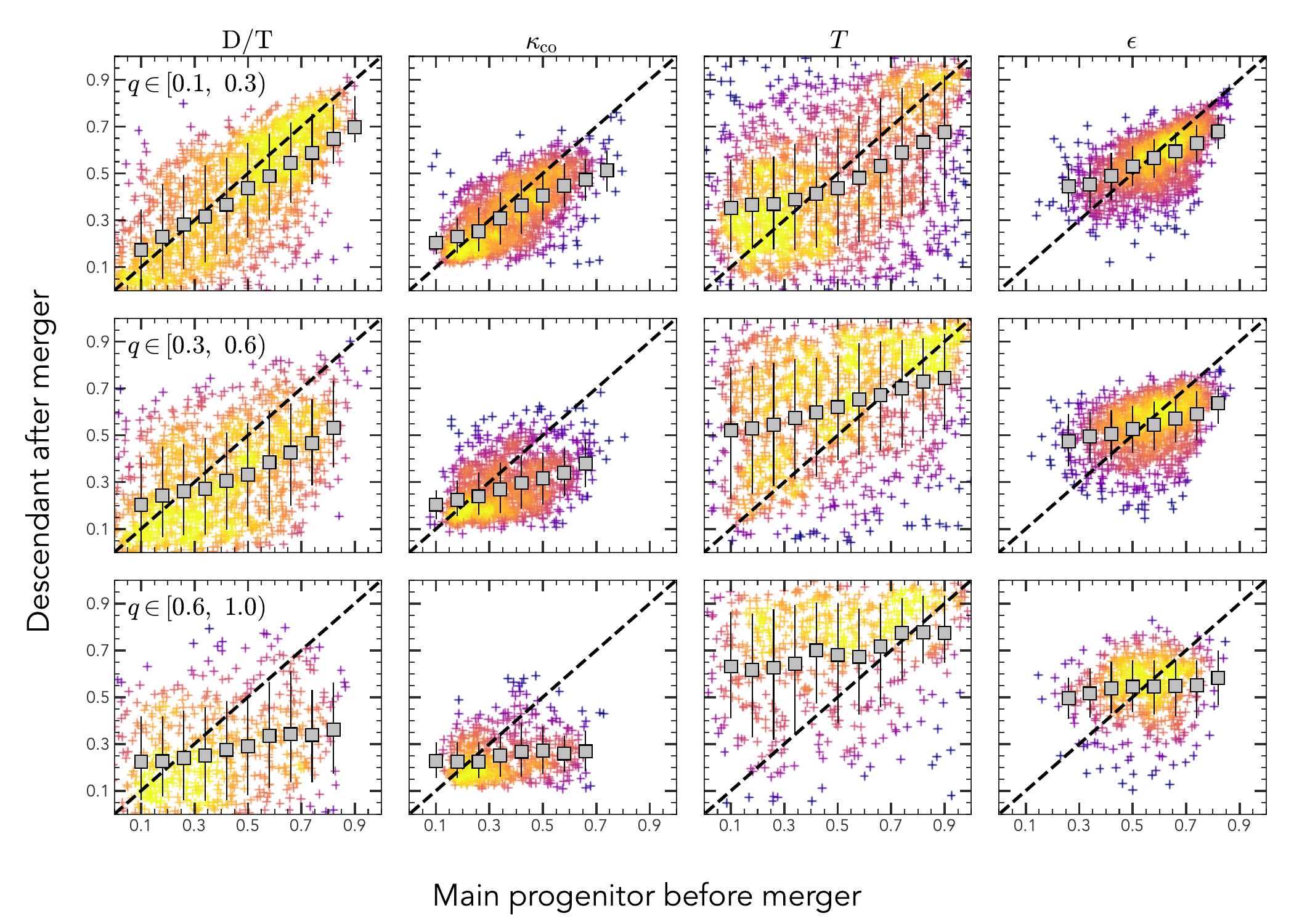}
  \end{center}
  \caption{
    Morphological and kinematical transformation before and after mergers.
    Each panel compares a property of the main progenitor before merger ($x$-axis) to that of the descendant galaxy after the merger ($y$-axis).
    Rows show three merger mass-ratio bins, $q\equiv M_{\star, 2}/M_{\star, 1}$: $0.1\leq q < 0.3$ (top), $0.3\leq q < 0.6$ (middle), and $0.6\leq q < 1$ (bottom).
    Columns show, from left to right, disc-to-total stellar mass ratio ($\rm D/T$), co-rotation kinetic-energy fraction ($\kappa_{\rm co}$), triaxiality ($T$), and flattening ($\epsilon$).
    Crosses show individual merger events and are coloured by the local density of mergers in 0.1-wide bins.
    Grey squares give the median descendant values in bins of the progenitor values with error bars indicating the 16th--84th percentiles.
    The dashed line marks equality.
    Mergers reduce rotational support (lower $\rm D/T$ and $\kappa_{\rm co}$) and shift remnants toward more prolate morphologies (higher $T$), with the effect strengthening at higher $q$.
  }
  \label{fig:merger_morphology}
\end{figure*}

\begin{figure*}
  \begin{center}
    \includegraphics[width=0.95\linewidth]{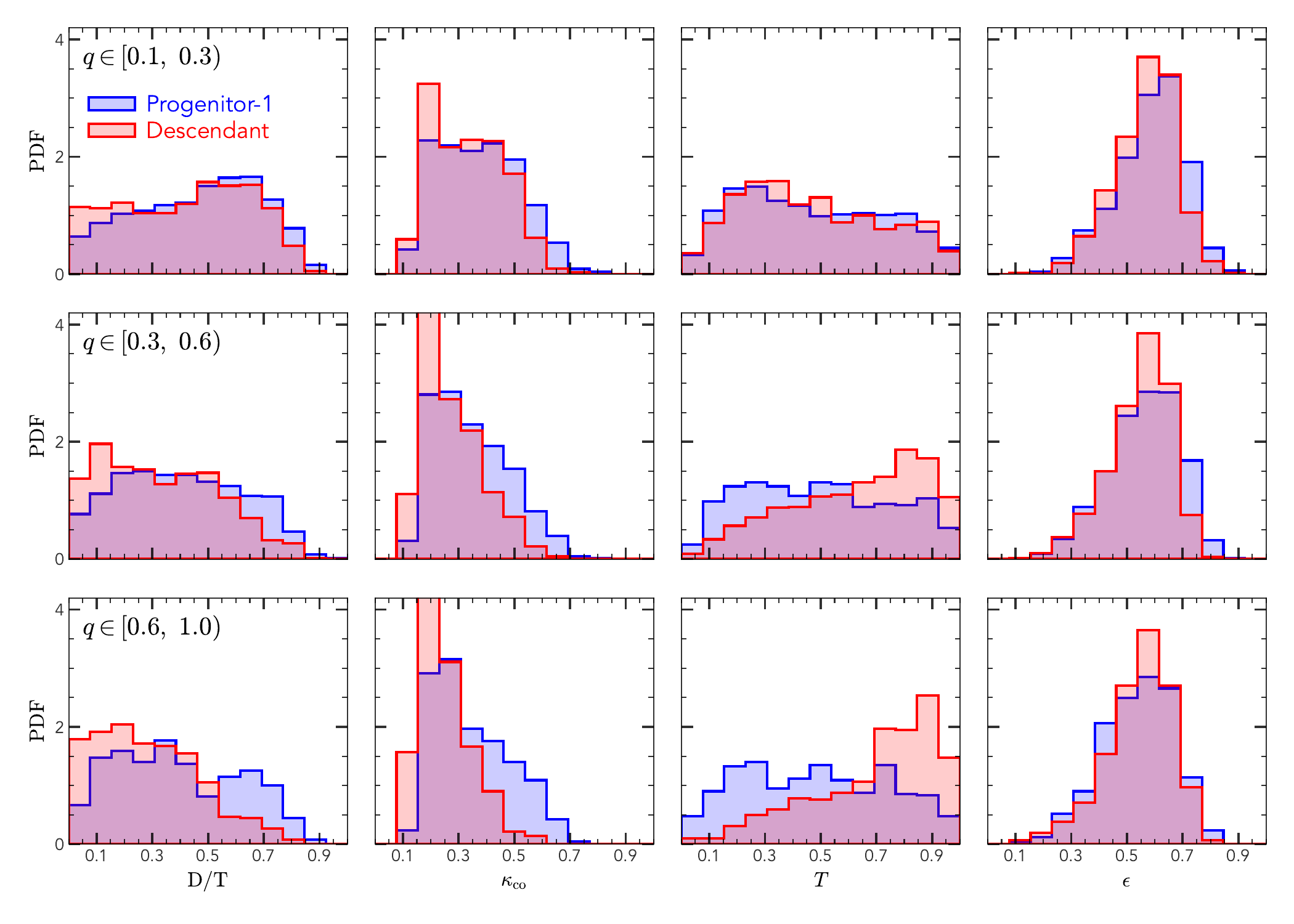}
  \end{center}
  \caption{
    Distributions of kinematical and morphological properties before and after mergers.
    Panels show the probability distribution functions of disc-to-total stellar mass ratio ($\rm D/T$), co-rotation kinetic-energy fraction ($\kappa_{\rm co}$), triaxiality ($T$), and flattening ($\epsilon$) of the main progenitor galaxy before merger (blue) and the descendant galaxy after merger (red).
    Rows correspond to merger mass-ratio bins, $q\equiv M_{\star, 2}/M_{\star, 1}$: $0.1\leq q < 0.3$ (top), $0.3\leq q < 0.6$ (middle), and $0.6\leq q < 1$ (bottom).
    Mergers reduce rotational support (lower $\rm D/T$ and $\kappa_{\rm co}$) and shift remnants toward more prolate morphologies (higher $T$), with the effect strengthening at higher $q$.
  }
  \label{fig:merger_hist}
\end{figure*}

Galaxy mergers redistribute stellar angular momentum and may increase random motions, thereby potentially lowering rotational support and building spheroids \citep[e.g.][]{naabStatisticalPropertiesCollisionless2003}.
Collisionless mergers drive violent relaxation and can produce triaxial remnants, and major mergers may also yield prolate remnants in both idealised and cosmological settings \citep[e.g.][]{ebrovaGalaxiesProlateRotation2017, liOriginPropertiesMassive2018}.

Figs.~\ref{fig:merger_morphology} and \ref{fig:merger_hist} quantify these trends in EAGLE using the diagnostics of \citet{thobRelationshipMorphologyKinematics2019}.
Fig.~\ref{fig:merger_morphology} shows that minor mergers ($0.1 \leq q < 0.3$) perturb galaxies but approximately preserve their rotational support: descendant galaxies scatter around the one-to-one relation in $\rm D/T$ and $\kappa_{\rm co}$ with a small median shift to lower values.
This shift strengthens with increasing mass ratio.
This behaviour matches the picture from controlled experiments in which minor mergers primarily thicken discs and heat stellar orbits without fully erasing rotation \citep{coxEffectGalaxyMass2008}.
For mergers with $0.6\leq q < 1$, descendant galaxies lie predominantly below the one-to-one line in both $\rm D/T$ and $\kappa_{\rm co}$, consistent with efficient conversion of ordered rotation into random motion during violent relaxation.
The median offset is largest for progenitors with high $\rm D/T$ and $\kappa_{\rm co}$, indicating that mergers most strongly transform disc-dominated systems.
Appendix~\ref{sec:the_dependence_on_the_main_progenitor_stellar_mass} shows that the morphological and kinematical transformation of galaxies depends weakly, if at all, on the main progenitor stellar mass.

Importantly, however, even the most nearly equal-mass mergers in our sample do not always fully destroy the disc.
Some descendant galaxies retain residual rotational support across the full range of progenitor D/T and $\kappa_{\rm co}$, and the post-merger distributions in Fig.~\ref{fig:merger_hist} retain a tail extending to high D/T and $\kappa_{\rm co}$ even for mergers with $q > 0.6$.
This is in tension with the assumption adopted in many semi-analytic models, where major mergers are assumed to completely destroy the disc and transfer all stellar mass into a spheroidal component \citep[e.g.][]{coleHierarchicalGalaxyFormation2000, guoDwarfSpheroidalsCD2011, laceyUnifiedMultiwavelengthModel2016}.
Our results suggest that this is an oversimplification: mergers drive substantial but incomplete morphological transformation, and a residual disc component typically survives even after a major merger \citep[see also][]{springelFormationSpiralGalaxy2005, kannanDiscsBulgesEffect2015, zengFormationMassiveDisc2021}.
This has implications for the predicted disc fraction and the build-up of spheroids in semi-analytic models.

Mergers also increase triaxiality.
Descendant galaxies tend to move above the one-to-one relation in $T$, and the effect grows with merger mass ratio, $q$.
Many mergers with high $q$ approach $T\sim 1$, signalling a shift toward prolate-like shapes, in line with merger remnants in collisionless simulations \citep[see also][]{naabStatisticalPropertiesCollisionless2003} and with the prolate populations reported in cosmological simulations \citep[see also][]{ebrovaGalaxiesProlateRotation2017, liOriginPropertiesMassive2018, lagosDiverseNatureFormation2022}.
Changes in flattening, $\epsilon$, are weaker and show substantial scatter at fixed progenitor $\epsilon$, suggesting that the merger primarily reshuffles the orbital structure rather than imposing a unique flattening.

Fig.~\ref{fig:merger_hist} shows the corresponding population-level shifts.
The post-merger distributions move to lower $\rm D/T$ and $\kappa_{\rm co}$, with the largest change for mergers with the highest $q$.
The triaxiality ($T$) distributions develop an enhanced high-$T$ peak after mergers, increasing the fraction of prolate remnants.
The flattening ($\epsilon$) distributions overlap more closely before and after merger.

\section{Summary}
\label{sec:summary}

We have used the EAGLE cosmological hydrodynamical simulation to investigate the impact of mergers on size-growth and morphological evolution of galaxies, focusing on post-merger galaxy size and structural transformation.
Our main conclusions are as follows:

\begin{enumerate}

  \item
        The energy-conservation post-merger size estimator of \citet{coleHierarchicalGalaxyFormation2000} in equation~\eqref{eq:merger_size} performs well in a cosmological setting.
        The predicted galaxy size reproduces the simulated descendant galaxy size across all merger mass ratio bins, except a moderate over-prediction for mergers with compact remnants.
        Spearman's rank correlation coefficient between the estimated galaxy size and the simulated descendant size is $\mathcal R\approx 0.8$ and the residual scatter is about $0.12$--$0.15$ dex, with an outlier fraction $\lesssim 5$ per cent (see Fig.~\ref{fig:merger_size}).
        These results validate the use of this estimator in semi-analytic models of galaxy formation.

  \item
        The dissipation correction of \citet{hopkinsDissipationExtraLight2009}, calibrated on idealised simulations, over-corrects for gas-rich mergers by up to $\approx 0.4$ dex and increases the overall scatter to $\sigma \approx 0.17$--$0.19$~dex (see Fig.~\ref{fig:merger_size_fgasbin}).
        A recalibration of the dissipation correction against the residuals in EAGLE yields only marginal improvement, and we therefore retain the original estimator throughout.

  \item
        Mergers drive systematic size growth.
        The per-merger logarithmic size growth ranges from $\lesssim 0.03$~dex for minor mergers ($q \approx 0.1$) to $\sim 0.10$~dex for equal-mass mergers ($q \approx 1$), and is well reproduced by the energy-conservation estimator across the full range of $q$ (see Fig.~\ref{fig:postmerger_size_inc}).
        The size growth efficiency per unit accreted mass, $\eta \equiv \mathrm{d}\log_{10} r_{\star}/\mathrm{d}\log_{10} M_{\star}$, spans $0.4$--$1.0$ for major mergers and rises steeply towards $\eta \approx 2$ for purely collisionless minor mergers in the limit $f_{\rm orbit}/c \to 0$, which constitutes the absolute upper limit of merger-driven size growth (see Fig.~\ref{fig:merger_driven_size}).
        Comparing to observationally inferred values, the $\eta \approx 2$ reported by \citet{vandokkumGROWTHMASSIVEGALAXIES2010} saturates this upper limit and requires unrealistically favourable merger conditions, while the more moderate $\eta \approx 1.4$ of \citet{hillMassColorStructural2017} still demands assembly predominantly through minor mergers with a small orbital energy contribution ($f_{\rm orbit}/c \lesssim 0.5$).
        The size growth efficiency $\eta$ is highly sensitive to the orbital energy at the time of merging, implying that the minor merger channel cannot be established as the driver of the rapid size growth of massive galaxies without better constraints on this quantity.

  \item
        Mergers systematically reduce rotational support and increase triaxiality.
        Descendants have lower D/T and $\kappa_{\rm co}$ than their progenitors, and the magnitude of this shift grows with the merger progenitor stellar mass ratio, $q$.
        Minor mergers ($0.1\leq q< 0.3$) perturb galaxies but largely preserve the morphology and kinematics of the main progenitor.
        Major mergers ($q > 0.3$) drive efficient conversion of ordered rotation into random motion, with disc-dominated progenitors experiencing the strongest transformation.
        Mergers also increase triaxiality, with the highest mass ratio mergers producing a significant population of prolate-like remnants (see Fig.~\ref{fig:merger_morphology}).
        Importantly, even the most nearly equal-mass mergers do not always fully destroy the disc, in tension with the complete disc destruction assumed in several semi-analytic models.
        These trends persist at the population level (see Fig.~\ref{fig:merger_hist}) and are insensitive to the main progenitor stellar mass (see Fig.~\ref{fig:merger_morphology_mstarbin}).

\end{enumerate}

Together, these results provide a simulation-validated framework for quantifying merger-driven size growth and structural transformation in a cosmological context.
The energy-conservation estimator is shown to be reliable across a wide range of merger mass ratios and progenitor properties, supporting its use in semi-analytic models of galaxy formation.
The sensitivity of $\eta$ to the orbital energy at the time of merging demonstrates that the role of minor mergers in driving the rapid size growth of massive galaxies cannot be established without better constraints on this quantity, motivating a more complete treatment of orbital energy in the size growth prescription of semi-analytic models.
The finding that even major mergers do not always fully destroy the disc further suggests that the morphological transformation prescriptions currently employed in semi-analytic models warrant revision.
We stress, however, that these conclusions are derived from the EAGLE cosmological hydrodynamical simulation alone, and it remains to be tested whether the predicted size growth efficiencies and morphological transformation rates are consistent with observational constraints from galaxy population statistics.

\section*{Acknowledgements}

KW thanks the inspiring discussion with Yong Shi, Andrew Pontzen, Joop Schaye, Aaron Ludlow, Katy L. Proctor, Enci Wang, Shude Mao, and Yangyao Chen at different stages of this work.
KW acknowledges the use of Claude (Anthropic) as a writing and research aid in the preparation of this manuscript, including literature searches, drafting assistance, and language editing; all scientific content, analysis, and conclusions are the authors' own.

This work is supported by the Science and Technology Facilities Council (STFC) through grant ST/X001075/1.
SB is supported by the UK Research and Innovation (UKRI) Future Leaders Fellowship [grant number MR/V023381/1 and UKRI2044].

This work is co-funded by the European Union (Widening Participation, ExGal-Twin, GA 101158446). Views and opinions expressed are however those of the author(s) only and do not necessarily reflect those of the European Union. Neither the European Union nor the granting authority can be held responsible for them.

This work used the DiRAC@Durham facility managed by the Institute for Computational Cosmology on behalf of the STFC DiRAC HPC Facility
(www.dirac.ac.uk).
The equipment was funded by BEIS capital funding via STFC capital grants ST/K00042X/1, ST/P002293/1, ST/R002371/1 and ST/S002502/1, Durham University and STFC operations grant ST/R000832/1.
DiRAC is part of the National e-Infrastructure.

\section*{Data availability}

The data underlying this article will be shared on reasonable request to the corresponding author.

\bibliographystyle{mnras}
\bibliography{bibtex.bib}

\appendix

\section{Sensitivity to the orbital energy parameter}
\label{sec:Optimising_forbit}

\begin{figure}
  \begin{center}
    \includegraphics[width=0.95\linewidth]{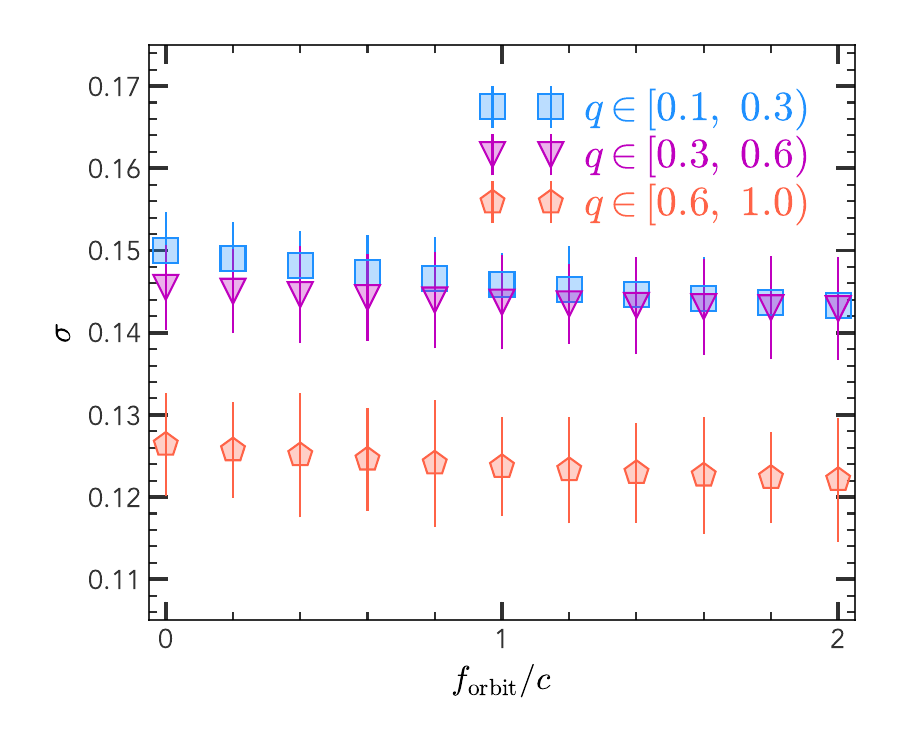}
  \end{center}
  \caption{
    Residual scatter $\sigma$ of the post-merger size estimator as a function of the orbital energy parameter $f_{\rm orbit}/c$, for three bins of merger mass ratio $q \equiv M_{\star,2}/M_{\star,1}$.
    Error bars indicate the uncertainty on $\sigma$ estimated from 100 bootstrap realisations.
    The scatter decreases weakly and monotonically with $f_{\rm orbit}/c$ in all three bins, by an amount comparable to the bootstrap uncertainty.
  }
  \label{fig:optimise_f_orbit.pdf}
\end{figure}

\begin{figure}
  \begin{center}
    \includegraphics[width=0.95\linewidth]{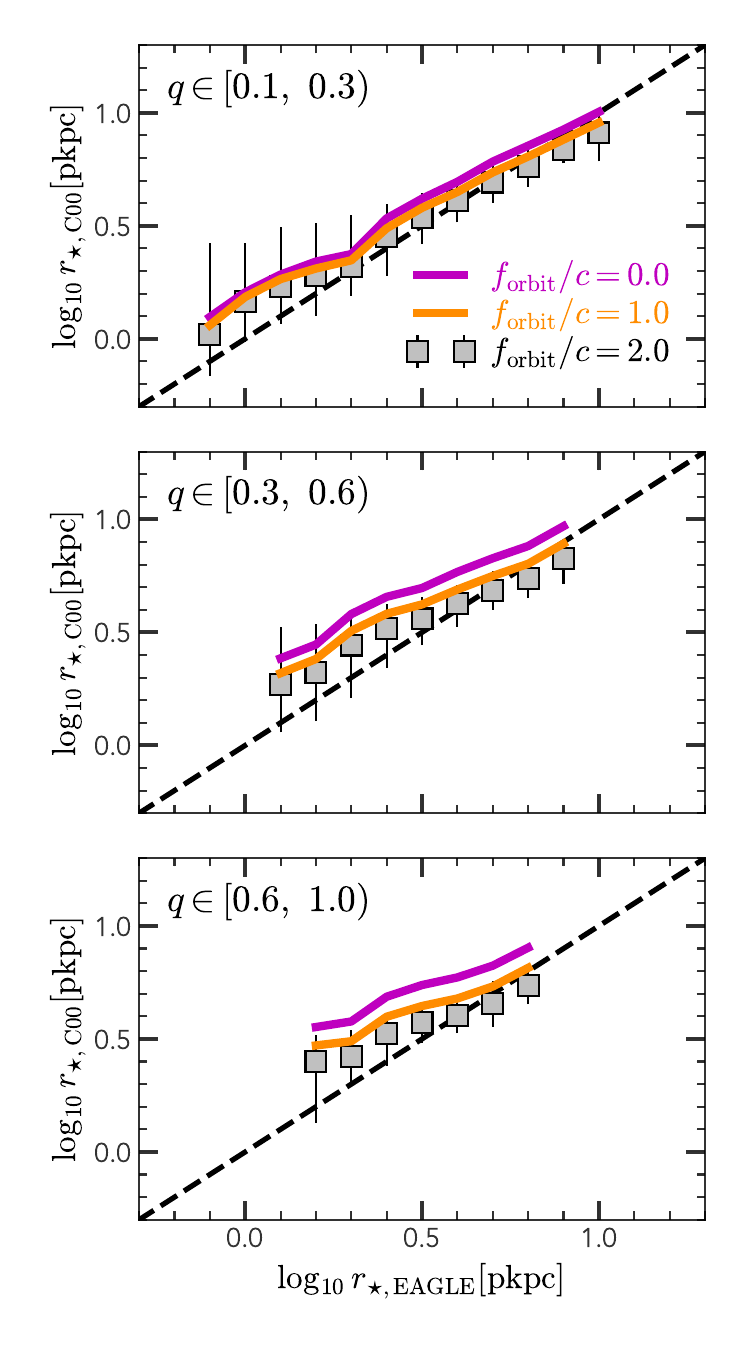}
  \end{center}
  \caption{
    Similar to Fig.~\ref{fig:merger_size}, except that the magenta and orange solid lines show the result for $f_{\rm orbit}/c$ = 0 and 1, respectively.
  }
  \label{fig:merger_size_diff_fo.pdf}
\end{figure}

The orbital energy parameter $f_{\rm orbit}/c$ in equation~\eqref{eq:merger_size} controls the contribution of the encounter energy to the remnant binding energy and is not directly accessible from the progenitor properties alone.
In the main text we adopt $f_{\rm orbit}/c = 2$ following \citet{coleHierarchicalGalaxyFormation2000}, corresponding to two point masses on a circular orbit separated by $r_{\star, 1} + r_{\star, 2}$.
Here we assess the sensitivity of the estimator to this choice by scanning $f_{\rm orbit}/c$ over the range $[0, 2]$ and measuring both the residual scatter $\sigma$ and the bias of the predicted remnant sizes in each merger mass-ratio bin.

Fig.~\ref{fig:optimise_f_orbit.pdf} shows the residual scatter $\sigma$ of $\log_{10}(r_{\star, \rm C00}/r_{\star, \rm EAGLE})$.
In all three mass-ratio bins $\sigma$ decreases weakly and monotonically with $f_{\rm orbit}/c$, by less than $0.01$~dex over the full range, which is comparable to the bootstrap uncertainty.
Mergers with $q\in [0.1, 0.6)$ remain at $\sigma \approx 0.14$--$0.15$~dex and mergers with $q > 0.6$ at $\sigma \approx 0.12$--$0.13$~dex.
The scatter alone therefore does not distinguish between values of $f_{\rm orbit}/c$ across this range.

Fig.~\ref{fig:merger_size_diff_fo.pdf} compares the median predicted size to the simulated descendant size for $f_{\rm orbit}/c = 0$, $1$ and $2$.
Smaller values over-predict the descendant size systematically, with the bias growing with galaxy size and becoming most pronounced for major mergers, while $f_{\rm orbit}/c = 2$ traces the one-to-one relation most closely.
The median relation therefore motivates our choice of $f_{\rm orbit}/c = 2$, which we retain throughout the paper.

\section{Dependence of the size estimation residual on progenitor galaxy properties}
\label{sec:size_residual_dependence}

\begin{figure*}
  \begin{center}
    \includegraphics[width=0.95\linewidth]{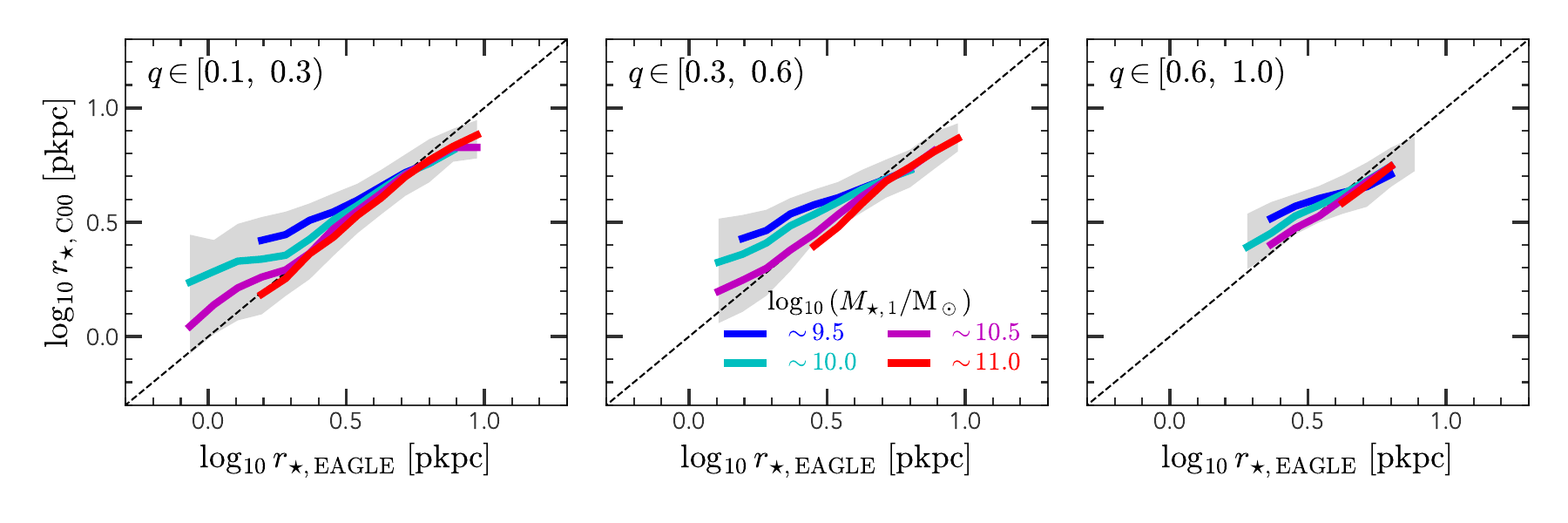}
  \end{center}
  \caption{
  Post-merger galaxy size predicted by the energy-conservation estimator compared to the simulated descendant size in EAGLE, in bins of merger mass ratio (panels) and main progenitor stellar mass $\log_{10} M_{\star,1}/{\rm M}_\odot$: ${\sim}9.5$, ${\sim}10.0$, ${\sim}10.5$, and ${\sim}11.0$, each with a bin width of 0.3~dex.
  The grey shaded region shows the 16th--84th percentile range of the full population.
  The median relations in different stellar mass bins trace each other closely across all merger mass-ratio bins, demonstrating that the estimator performance has no systematic dependence on the main progenitor stellar mass.
  }
  \label{fig:merger_size_mstarbin}
\end{figure*}

\begin{figure*}
  \begin{center}
    \includegraphics[width=0.95\linewidth]{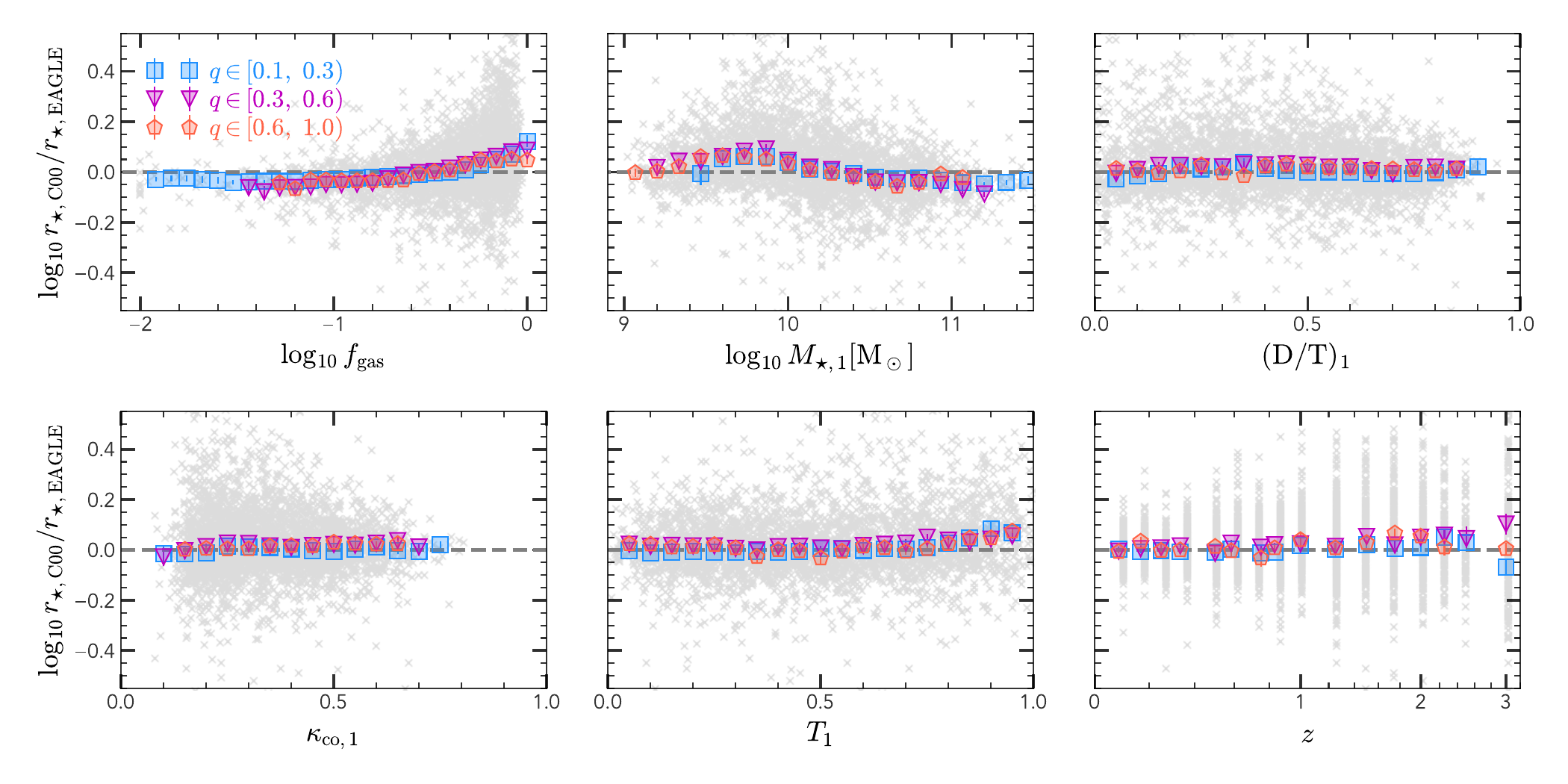}
  \end{center}
  \caption{
    Residual of the post-merger galaxy size estimate, $\log_{10}(r_{\star,\rm C00}/r_{\star,\rm EAGLE})$, as a function of progenitor properties in bins of merger mass ratio (colours).
    Panels show, from left to right and top to bottom, dependence on the progenitor gas fraction $f_{\rm gas}$, main progenitor stellar mass $M_{\star,1}$, main progenitor disc-to-total stellar mass ratio $\rm D/T$, main progenitor co-rotation fraction $\kappa_{\rm co,1}$, main progenitor triaxiality $T_1$, and redshift, respectively.
    The median residual remains close to zero with only weak dependence on progenitor properties across all merger mass-ratio bins.
  }
  \label{fig:merger_size_residual}
\end{figure*}

Fig.~\ref{fig:merger_size_mstarbin} shows the predicted versus simulated descendant size in bins of main progenitor stellar mass.
The median relations in different stellar mass bins trace each other closely across all merger mass-ratio bins, with no systematic offset indicating a mass-dependent estimator performance.
This confirms that the good agreement reported in Section~\ref{sec:predicting_the_size_of_merger_remnant} is not restricted to a particular mass range but holds across the full stellar mass range probed here.

Fig.~\ref{fig:merger_size_residual} shows the residual $\log_{10}(r_{\star, \rm C00}/r_{\star,\rm EAGLE})$ as a function of six progenitor properties.
The median residual is close to zero across all panels and merger mass-ratio bins, confirming the absence of any strong systematic bias.
The one exception occurs at high gas fractions ($f_{\rm gas} \sim 1$), where the estimator over-predicts the descendant size by $\approx 0.1$~dex for minor mergers ($q \in [0.1, 0.3)$).
This is consistent with the expectation that gas-rich mergers are more dissipative \citep{coxKinematicStructureMerger2006, covingtonRoleDissipationScaling2011, moTwophaseModelGalaxy2024, chenTwophaseModelGalaxy2024a, chenTwophaseModelGalaxy2025}: the inflow of gas towards the galaxy centre and the resulting nuclear starburst produce a more compact remnant than the purely collisionless energy-conservation argument predicts, and this effect is most pronounced for minor mergers where the secondary contributes a larger gas reservoir relative to its stellar mass.
The major merger ($q > 0.3$) bins show no comparable offset at high gas fractions, suggesting that the dissipative correction is subdominant when the mass ratio is large enough for violent relaxation to dominate the structural rearrangement.
The residual shows no measurable dependence on $M_{\star,1}$ across $10^9$--$10^{11}\,\rm M_\odot$, nor on the main progenitor morphological and kinematical properties $({\rm D/T})_1$, $\kappa_{\rm co,1}$, and $T_1$, and no trend with redshift is detected over $0 \leq z \leq 3$.
Together with Fig.~\ref{fig:merger_size_mstarbin}, these results demonstrate that the estimator is robust to the diversity of progenitor structures and cosmic epochs sampled by the EAGLE merger population.

\section{Information gain from including the secondary progenitor size}
\label{sec:info_gain}

\begin{figure}
  \begin{center}
    \includegraphics[width=0.95\linewidth]{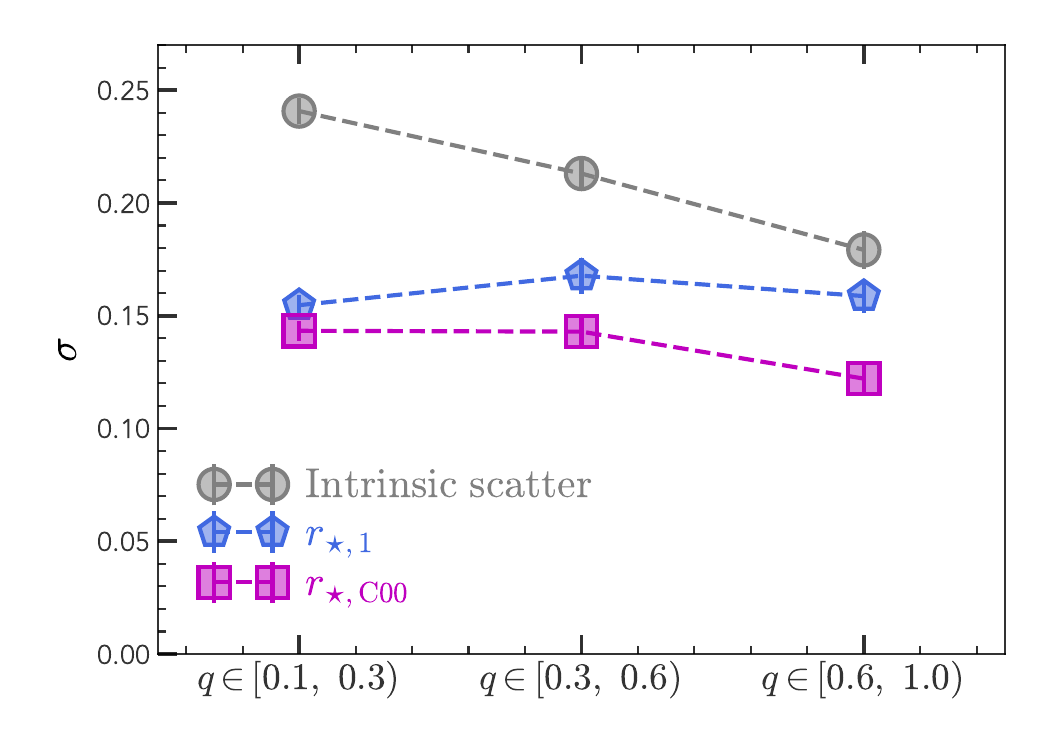}
  \end{center}
  \caption{
    Residual scatter $\sigma$ of three predictors of the descendant galaxy size as a function of merger mass ratio bin.
    Grey circles show the intrinsic scatter of the descendant size $r_{\star,\rm EAGLE}$ about the mean size--mass relation, representing the baseline scatter in the absence of any merger information.
    Blue pentagons show the scatter when the main progenitor size $r_{\star, 1}$ alone is used as the predictor.
    Magenta squares show the scatter of the energy-conservation estimator $r_{\star,\rm C00}$, which incorporates both progenitor sizes and the secondary progenitor mass.
    Error bars indicate bootstrap uncertainties from 100 realisations.
  }
  \label{fig:extra_mp.pdf}
\end{figure}

The energy-conservation estimator in equation~\eqref{eq:merger_size} requires the size of the secondary progenitor, $r_{\star, 2}$, in addition to the main progenitor size $r_{\star, 1}$ and the mass ratio $q$.
In semi-analytic models, $r_{\star,2}$ is available, and here we quantify how much predictive power it contributes relative to using $r_{\star,1}$ alone as a proxy for the descendant size.

Fig.~\ref{fig:extra_mp.pdf} compares three levels of prediction across the three merger mass-ratio bins.
The intrinsic scatter of the descendant size about the mean size--mass relation (grey) decreases modestly from $\sigma \approx 0.24$~dex for minor mergers to  $\approx 0.18$~dex for major mergers, reflecting the greater structural diversity of remnants at low mass ratio.
Using $r_{\star, 1}$ alone already reduces the scatter substantially, to $\sigma \approx 0.16$~dex across all bins.
The full energy-conservation estimator $r_{\star,\rm C00}$ further reduces the scatter to $\sigma \approx 0.12$--$0.15$~dex for merger with $q \in [0.1, 0.6)$, and to $\approx 0.12$~dex for $q > 0.6$.

The improvement from including secondary progenitor information is therefore modest for minor mergers, where the secondary contributes little mass and $r_{\star,\rm C00}$ is dominated by $r_{\star, 1}$, but becomes more meaningful for most nearly equal-mass mergers ($q \in [0.6, 1.0)$), where the two progenitors are comparable in mass and the secondary size and orbital energy term together reduce the residual scatter by $\approx 0.03$~dex relative to using $r_{\star, 1}$ alone.
These results confirm that the secondary progenitor carries genuine additional information about the remnant size, and that its inclusion in the estimator is most valuable in the major merger regime.

\section{Derivation of the analytic size growth formula}
\label{sec:derivation_size_growth}

Starting from the energy-conservation estimator of equation~(\ref{eq:merger_size})
\begin{equation}
  \frac{(M_{\star,1} + M_{\star,2})^2}{r_{\star,\rm C00}} =
  \frac{M_{\star,1}^2}{r_{\star,1}} +
  \frac{M_{\star,2}^2}{r_{\star,2}} +
  \frac{f_{\rm orbit}}{c}
  \frac{M_{\star,1} M_{\star,2}}{r_{\star,1} + r_{\star,2}},
  \label{eq:merger_size_appendix}
\end{equation}
we divide both sides by $M_{\star,1}^2/r_{\star,1}$ and introduce the stellar mass ratio $q \equiv M_{\star,2}/M_{\star,1}$.
Assuming that the progenitor galaxies follow a power-law mass--size relation $r_{\star} \propto M_{\star}^p$, the secondary progenitor size can be written as $r_{\star,2} = q^p r_{\star,1}$.
Substituting these into equation~\eqref{eq:merger_size_appendix} gives
\begin{equation}
  \frac{(1 + q)^2}{r_{\star,\rm C00}/r_{\star,1}} =
  1 + q^{2-p} +
  \frac{f_{\rm orbit}}{c}
  \frac{q}{1 + q^p},
\end{equation}
where we have used $r_{\star,1} + r_{\star,2} = r_{\star,1}(1 + q^p)$ and $M_{\star,1}M_{\star,2}/(M_{\star,1}^2/r_{\star,1}) = q\,r_{\star,1}$.
Rearranging gives the predicted size growth per merger event,
\begin{equation}
  \frac{r_{\star,\rm C00}}{r_{\star,1}} = (1+q)^2
  \left(1 + q^{2-p} + \frac{f_{\rm orbit}}{c}
  \frac{q}{1+q^p}\right)^{-1}.
  \label{eq:size_growth_appendix}
\end{equation}
In the limits of vanishing and equal-mass mergers, this reduces to
\begin{align}
  \frac{r_{\star,\rm C00}}{r_{\star,1}}\bigg|_{q\to 0} & = 1, \\
  \frac{r_{\star,\rm C00}}{r_{\star,1}}\bigg|_{q=1}    & =
  \frac{4}{2 + (f_{\rm orbit}/c)/2},
\end{align}
recovering the expected result that a vanishingly small merger leaves the remnant size unchanged, while an equal-mass merger with $f_{\rm orbit}/c = 2$ gives $r_{\star,\rm C00}/r_{\star,1} = 4/3$, corresponding to $\approx 0.12$~dex size growth.

The size growth efficiency per unit accreted stellar mass, $\eta$, follows directly from equation~\eqref{eq:size_growth_appendix},
\begin{align}
  \eta(p, q, f_{\rm orbit}/c) & \equiv
  \frac{\mathrm{d}\log_{10} r_\star}{\mathrm{d}\log_{10} M_\star}
  = \frac{\log_{10}(r_{\star,\rm C00}/r_{\star,1})}{\log_{10}(1+q)}
  \nonumber                                                                \\
                              & = 2 - \frac{\log_{10}\!\left(1 + q^{2-p} +
    \dfrac{f_{\rm orbit}}{c}\dfrac{q}{1+q^p}\right)}
  {\log_{10}(1+q)}.
  \label{eq:eta_appendix}
\end{align}
In the limits of vanishing and equal-mass mergers, this reduces to
\begin{align}
  \eta(p, q=0, f_{\rm orbit}/c) & = 2 - \frac{f_{\rm orbit}}{c},                                                    \\
  \eta(p, q=1, f_{\rm orbit}/c) & = 2 - \frac{\log_{10}\!\left(2 + \dfrac{f_{\rm orbit}}{2c}\right)} {\log_{10} 2},
\end{align}
showing that, for $p < 1$, $\eta$ is bounded above by 2 in the minor merger limit with zero orbital energy ($f_{\rm orbit}/c = 0$), and falls monotonically as $f_{\rm orbit}/c$ increases or as the mass ratio $q$ increases toward unity.
For $p = 1$ the minor merger limit is $\eta \rightarrow 1 - f_{\rm orbit}/c$ instead.

\section{Mass--size relation for merging progenitor galaxies}
\label{sec:mass_size_relation_for_merging_progenitor_galaxies}

\begin{figure}
  \begin{center}
    \includegraphics[width=0.95\linewidth]{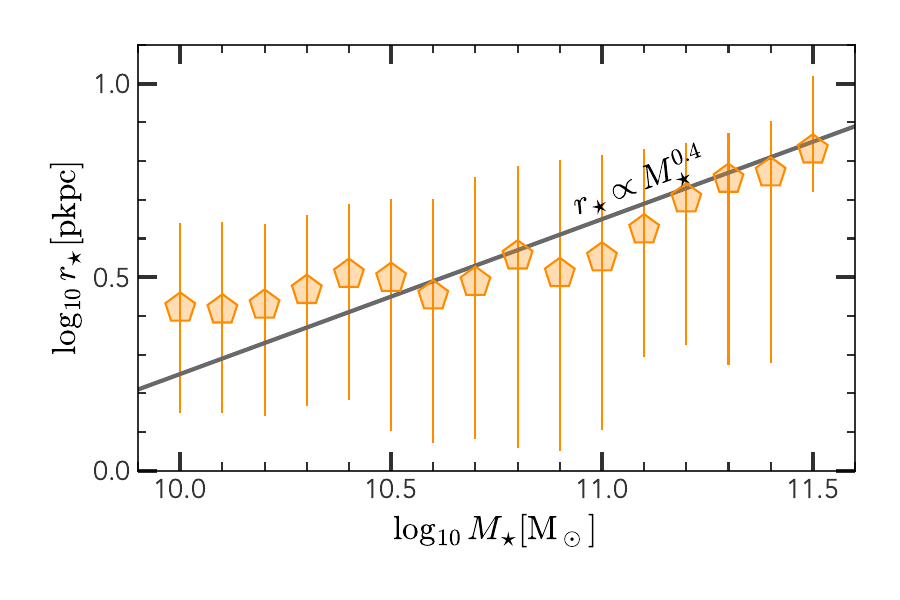}
  \end{center}
  \caption{
    The stellar mass--size relation for the main progenitor galaxies of our merger sample.
    Orange pentagons show the median $\log_{10} r_{\star}$ in bins of $\log_{10} M_{\star}$ of width 0.1~dex, with error bars indicating the 16th--84th percentile range.
    The dashed line shows a reference slope of $r_{\star} \propto M_{\star}^{0.4}$, and is not a fit to the data.
  }
  \label{fig:mass_size}
\end{figure}

The analytic size growth formula in equation~\eqref{eq:size_growth} assumes that the progenitor galaxies follow a power-law mass--size relation of the form $r_{\star} \propto M_{\star}^p$.
Fig.~\ref{fig:mass_size} shows the mass--size relation of the main progenitor galaxies in our merger sample, spanning the stellar mass range $10^{10}$--$10^{11.5}\,\rm M_\odot$.

The median size increases steadily with stellar mass over this range, broadly consistent with a power-law slope of $p \approx 0.4$ as indicated by the reference line.
The scatter is large, with 16th--84th percentile ranges of $\sim 0.3$--$0.4$~dex at fixed mass, reflecting the diversity of morphological types and formation histories in the merger progenitor population.
The slope $p = 0.4$ adopted in equation~\eqref{eq:size_growth} and Fig.~\ref{fig:postmerger_size_inc} is therefore a reasonable characterisation of the progenitor population as a whole, and the sensitivity of the results to this choice is explored in Fig.~\ref{fig:merger_driven_size}, where we show that varying $p$ between 0.2 and 0.6 has only a modest effect on $\eta$ across the full range of mass ratios.

\section{The dependence of morphology transformation on the main progenitor stellar mass}
\label{sec:the_dependence_on_the_main_progenitor_stellar_mass}

\begin{figure*}
  \begin{center}
    \includegraphics[width=0.95\linewidth]{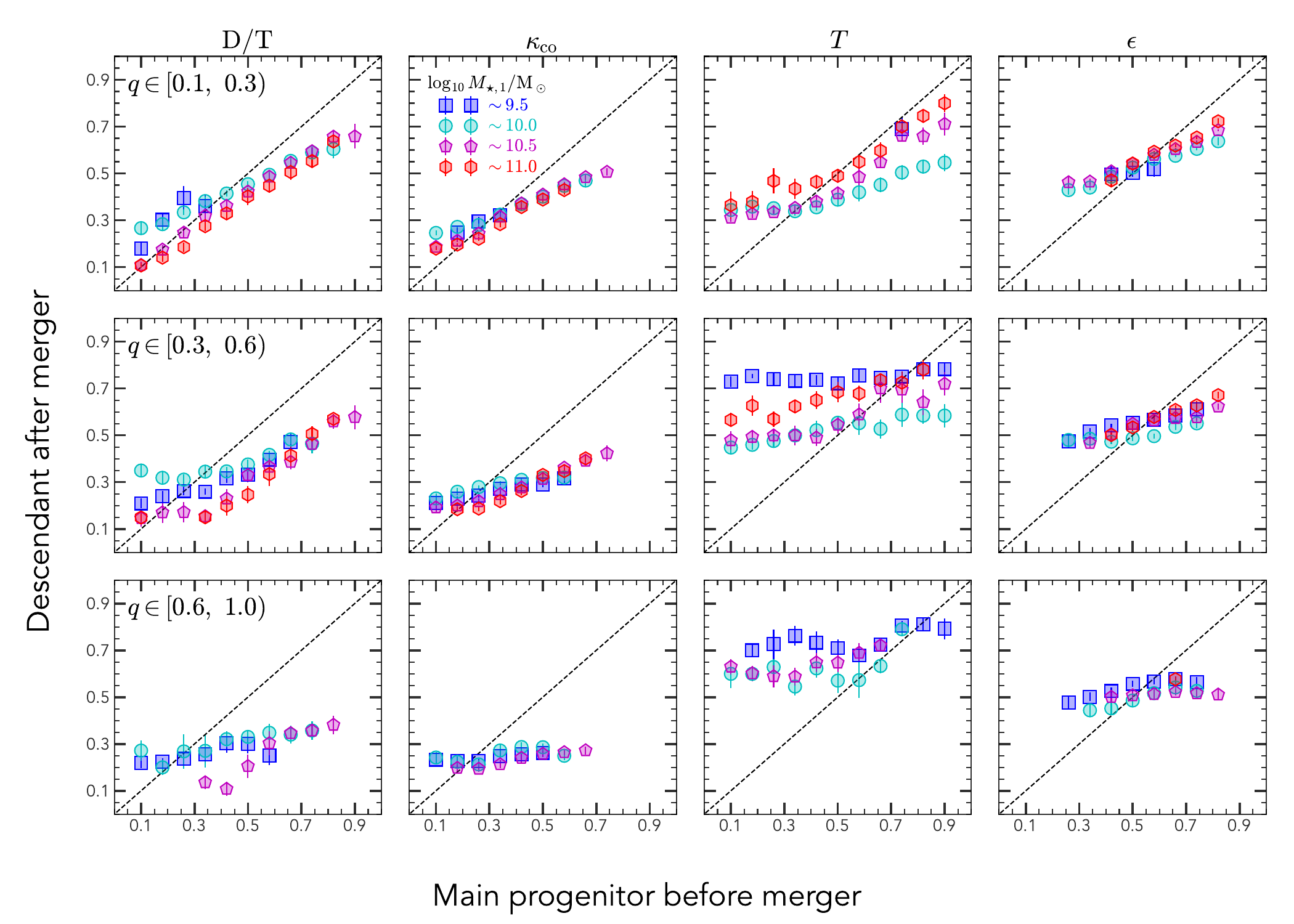}
  \end{center}
  \caption{
    Similar to Fig.~\ref{fig:merger_morphology}, except that we show the median relation in bins of the main progenitor stellar mass, $M_{\star, 1}$, with a bin width of 0.3 dex.
    Error bars give the standard deviation of the median from 100 bootstrap realisations.
    The morphological and kinematical transformation driven by mergers shows no systematic dependence on the main progenitor stellar mass across the full range probed here.
  }
  \label{fig:merger_morphology_mstarbin}
\end{figure*}

Fig.~\ref{fig:merger_morphology_mstarbin} shows the pre- to post-merger relations in ${\rm D/T}$, $\kappa_{\rm co}$, $T$, and $\epsilon$, split by main progenitor stellar mass across four bins spanning $\log_{10}(M_{\star,1}/\rm M_\odot) \approx 9.5$--$11.0$.

Across all four morphological and kinematical diagnostics and all three merger mass-ratio bins, the median relations in different stellar mass bins trace each other closely, with no systematic offset that would indicate a mass-dependent transformation efficiency.
In particular, the downward shift in ${\rm D/T}$ and $\kappa_{\rm co}$ below the one-to-one line, which strengthens with increasing mass ratio $q$, is present at all stellar masses with comparable magnitude.
Similarly, the upward shift in triaxiality $T$ for major mergers shows no clear dependence on $M_{\star, 1}$.
The flattening $\epsilon$ shows the largest scatter between mass bins, but no coherent trend with stellar mass.

These results demonstrate that the merger-driven morphological transformation quantified in Section~\ref{sec:morphology_transformation_after_merger} is not significantly modulated by the absolute stellar mass of the main progenitor over the range $10^{9.5}$--$10^{11.0}\,\rm M_\odot$ probed by our sample, and that mass ratio $q$ is the dominant parameter controlling the degree of structural transformation.

\bsp  
\label{lastpage}
\end{document}